\documentclass[acmsmall]{zkayart}

\makeatletter
\def\mdseries@tt{m}
\makeatother

\acmYear{2020}
\acmMonth{9}
\startPage{0}

\setcopyright{none}

\usepackage[utf8]{inputenc}
\usepackage[T1]{fontenc}

\usepackage{booktabs}   %
\usepackage{subcaption} %
\usepackage{graphicx}
\usepackage{setspace}
\usepackage{pdflscape}

\usepackage[frozencache]{minted}
\definecolor{mintedbg}{HTML}{F8F9FA}
\setminted{autogobble,stripnl=false,tabsize=2,frame=lines,framesep=2mm,fontsize=\footnotesize}
\AtBeginEnvironment{minted}{%
  }
\def\zkay{zkay_lexer.py:ZkayLexer -x} %

\usepackage{pgfplots}
\pgfplotsset{compat=1.14}

\usepackage[capitalise]{cleveref}

\usepackage{pdfpages}

\usepackage{enumitem}
\setlist[itemize]{label=$\triangleright$,topsep=0pt,itemsep=4pt}
\setlist[enumerate]{label=(\roman*),topsep=0pt,itemsep=4pt}

\usepackage{ragged2e}

\usepackage{xspace}

\newcommand{\name}[0]{zkay v0.2\xspace}
\newcommand{\Name}[0]{Zkay v0.2\xspace}
\newcommand{\oldname}[0]{zkay v0.1\xspace}
\newcommand{\Oldname}[0]{Zkay v0.1\xspace}

\newcommand{\code}[1]{\texttt{#1}}

\begin{document}
\pagenumbering{gobble}

\title[zkay 0.2]{zkay v0.2: Practical Data Privacy\\ for Smart Contracts}         %
\subtitle{\setstretch{1.5}Technical Report}     %

\author{Nick Baumann}
\affiliation{
  \institution{ETH Zürich}            %
  \country{Switzerland}                    %
}

\author{Samuel Steffen}
\affiliation{
  \institution{ETH Zürich}            %
  \country{Switzerland}                    %
}

\author{Benjamin Bichsel}
\affiliation{
  \institution{ETH Zürich}            %
  \country{Switzerland}                    %
}

\author{Petar Tsankov}
\affiliation{
  \institution{ETH Zürich}            %
  \country{Switzerland}                    %
}

\author{Martin Vechev}
\affiliation{
  \institution{ETH Zürich}            %
  \country{Switzerland}                    %
}

\begin{abstract}
\centering\includegraphics[width=0.5\textwidth]{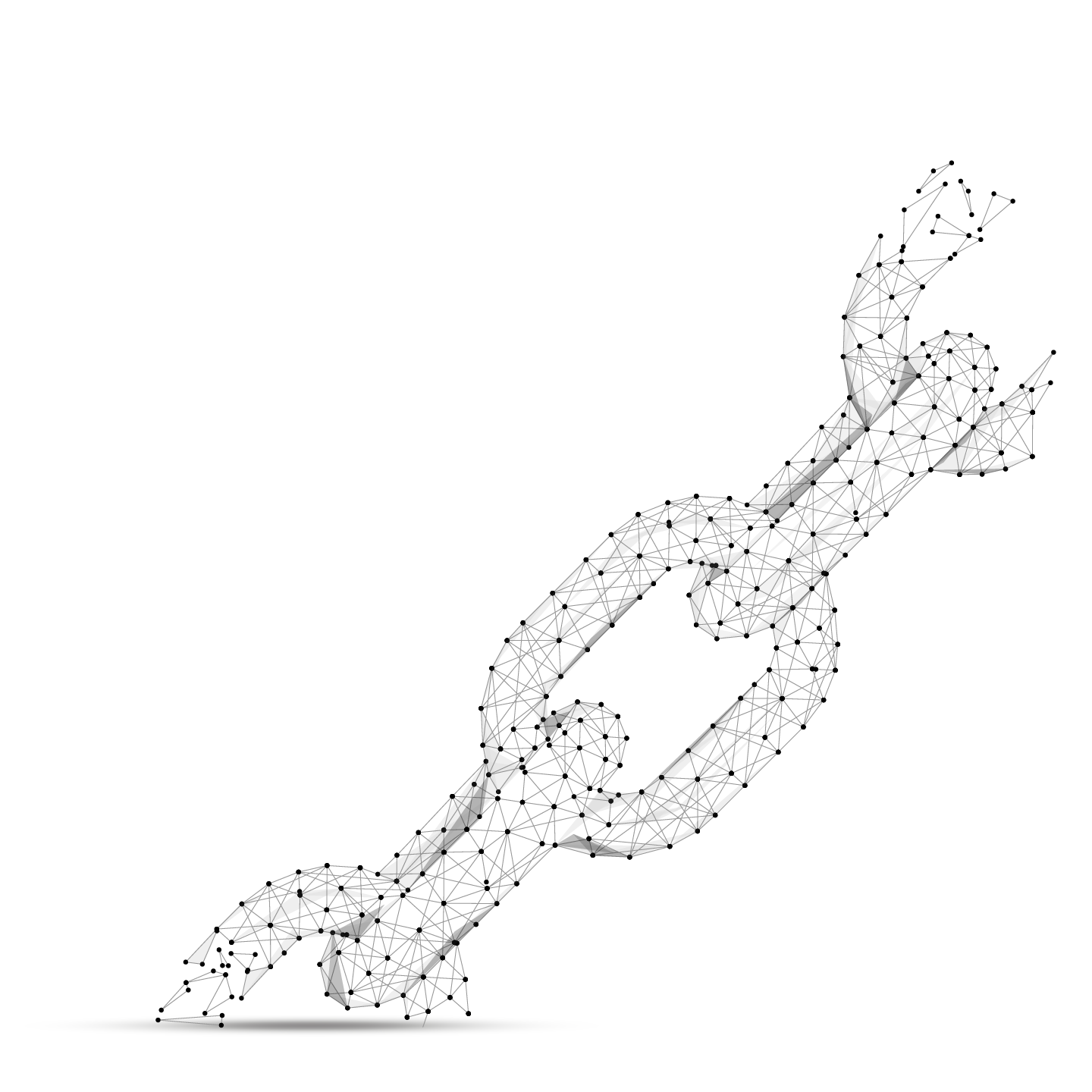}

\justifying
\clearpage
\section*{Abstract}
Recent work introduces zkay, a system for specifying and enforcing data privacy
in smart contracts. While the original prototype implementation of zkay (v0.1)
demonstrates the feasibility of the approach, its proof-of-concept
implementation suffers from severe limitations such as insecure encryption and
lack of important language features.

In this report, we present \name, which addresses its predecessor's limitations.
The new implementation significantly improves security, usability, modularity,
and performance of the system. In particular, zkay v0.2 supports
state-of-the-art asymmetric and hybrid encryption, introduces many new language
features (such as function calls, private control flow, and extended type
support), allows for different zk-SNARKs backends, and reduces both compilation
time and on-chain costs.

\end{abstract}

\topskip0pt
\vspace*{0.05\paperheight}
\maketitle

\vspace*{\fill}

\clearpage
\pagenumbering{arabic}
\section*{Contents}
\addcontentsline{toc}{section}{Contents}
\tableofcontents

\section*{Security Disclaimer}
\addcontentsline{toc}{section}{Security Disclaimer}

\Name is a research project and its implementation should \textbf{not} be
considered secure (e.g., it may contain bugs and has not undergone any security
review)! Do not use \name in a productive system or to process sensitive
confidential data. \Name is licensed under the MIT license and hence subject to
the following disclaimer.~\footnote{\url{https://opensource.org/licenses/MIT}}

\emph{The software is provided ``as is'', without warranty of any kind, express
or implied, including but not limited to the warranties of merchantability,
fitness for a particular purpose and noninfringement. In no event shall the
authors or copyright holders be liable for any claim, damages or other
liability, whether in an action of contract, tort or otherwise, arising from,
out of or in connection with the software or the use or other dealings in the
software.}

\section{Introduction}

\subsection{An Improved Implementation of zkay}

Zkay~\cite{zkay} is a system for specifying and enforcing data privacy in smart
contracts. The original publication comes with a prototype implementation of
zkay~\cite{zkay-poc}, which however suffers from various limitations. We refer
to this prototype implementation as \emph{\oldname}.

This technical report presents \emph{\name}~\cite{zkay-v2}, a significantly
improved implementation of zkay in terms of security, usability, modularity, and
performance. The report describes features and implementation details of \name,
and is targeted at both users and developers. It assumes that the reader is
familiar with the language and key ideas of zkay according to the original
publication~\cite{zkay}.

In summary, \name:

\begin{itemize}
	\item Adds support for state-of-the-art asymmetric and hybrid encryption;
	\item introduces essential language features including function calls,
	cryptocurrency-related functionality, private control flow, extended type
	support, and many more;
	\item supports different zk-SNARKs backends;
	\item provides a live transaction runtime automatically transforming
	transactions, computing the necessary zero-knowledge proofs, and directly
	interacting with a blockchain;
	\item features various usability improvements such as improved error
	messages and simplified installation; and
	\item reduces both compilation time and on-chain costs.
\end{itemize}

\subsection{Terminology}

We now introduce some terms which are commonly used in this report.

\begin{itemize}
	\item \emph{\Oldname}: The proof-of-concept implementation
	from~\cite{zkay-poc}.
	\item \emph{\Name}: The new version of zkay~\cite{zkay-v2} described in this
	report.
	\item \emph{Public computation/value}: An expression/value whose owner is
	\code{all}.
	\item \emph{Private computation/value}: An expression/value which is not
	public.
	\item \emph{Public function}: A function whose arguments and return values
	are public, and whose body only contains public computations and calls to
	public functions.  Note that unless explicitly stated otherwise, this does
	\emph{not} refer to the visibility modifier of the function.
	\item \emph{Private function}: A function which is not public.  Note that
	unless explicitly stated otherwise, this does \emph{not} refer to the
	visibility modifier of the function.
	\item \emph{Internal function call}: Calling a contract function from
	within the same contract.
	\item \emph{External function call}: Calling a contract function from an
	external account (i.e., a user or a different contract).
	\item \emph{On-chain}: Execution on the blockchain as part of a smart
	contract transaction. Increased on-chain computation results in higher
	Ethereum gas costs.
	\item \emph{Off-chain}: Local execution on a machine of a single user.
	\item \emph{Non-interactive zero-knowledge (NIZK) proof:} A proof of a
	statement about secret values which does not leak any information about
	these values besides their existence~\cite{nizk-orig,nizk-general}. The
	statement can be parameterized by public values. A NIZK proof is generated
	by a \emph{prover} knowing the secret values, and verified by a
	\emph{verifier}. In particular, NIZK proofs can be used to prove the correct
	execution of a computation  $out_{pub} = \phi(in_{pub}, in_{priv})$
	accepting public and private inputs $in_{pub}$ resp.\ $in_{priv}$, and
	producing a public result $out_{pub}$.
	\item \emph{Zk-SNARKs}: An efficient instantiation of
	NIZKs~\cite{zkSNARK}. Zk-SNARKS require a trusted setup phase, which
	generates a common reference string for proof generation and verification.
	In practice, most proving schemes generate a keypair $(k_p, k_v)$ from a
	constraint system representing the proof statement $\phi$ and secret
	randomness during the trusted setup phase. The keypair $(k_p, k_v)$ is
	publicly known: the \emph{prover key} $k_p$ is used to generate proofs for
	$\phi$ and arbitrary inputs, while the \emph{verification} key $k_v$ is
	required to verify such proofs.
	\item \emph{Abstract proof circuit}: A high-level, NIZK-framework agnostic
	representation of a proof statement (the computation $\phi$ to be proven)
	internally used by zkay. Such circuits are compiled to framework-specific
	concrete proof circuits (see below) by different backends.
	\item \emph{(Concrete) proof circuit}: An arithmetic prime field circuit
	expressing a proof statement. A circuit can have multiple input parameters,
	which can be private or public (see below). Zk-SNARK frameworks such as
	jsnark~\cite{jsnark-github} generate low-level constraint systems from such
	circuits, which are then used to generate NIZK key pairs and proofs.
	\item \emph{Private (circuit) input}: Input representing a secret to be
	hidden from the verifier. Private inputs are known to the prover.
	\item \emph{Public (circuit) input}: Input representing public knowledge,
	which is supplied to verifiers on the blockchain.
\end{itemize}

\section{Limitations of \oldname}

This section summarizes the main limitations of \oldname~\cite{zkay-poc}. The
key design goal of \name is to address these limitations.

\subsection{Security}\label{sec:nosec}

\paragraph{Insecure Encryption}
A major limitation of \oldname is its lack of secure encryption. In particular,
it uses the surrogate encryption function $\text{Enc}(v, k) = v + k$ to encrypt
plaintext $v$ with key $k$, which does not ensure confidentiality. \Oldname
leverages the ZoKrates framework~\cite{zokrates,zokrates-github} for NIZK proof
generation and verification, which at the time of development did not support
more realistic encryption functions.

\paragraph{Missing Integrity Verification}
\oldname does not provide tool support for verifying whether the bytecode of a
deployed smart contract corresponds to a given zkay contract. Doing this
manually is challenging due to the involvement of multiple contracts (main,
library, and verification contracts), and because exactly reproducing the
bytecode compilation output requires the same verification keys to be used.

\paragraph{Undefined Behavior}
The transaction transformation in \oldname uses arbitrary-precision integers
instead of emulating the under- and overflow semantics of fixed-sized integers.
The user is required to ensure the absence of under- and overflows as these may
result in security vulnerabilities such as the possibility of permanently
locking the funds of a contract.

Additionally, \oldname assumes that variables are always initialized and does
not model the zero-initialization semantics of Solidity. When reading an
uninitialized variable, transaction transformation in \oldname may therefore
produce an invalid proof.

\subsection{Language Fragment}

\oldname only supports a small subset of Solidity as indicated below. This
significantly restricts expressivity of zkay contracts.
\begin{itemize}
	\item only \code{bool}, \code{uint} (256-bit unsigned integer) and
	\code{mapping} types; no integer variant, \code{enum}, or private
	\code{address} types
	\item no user-defined types
	\item no tuples or multiple return values
	\item only basic statements (\code{require} and assignment); no control flow
	except \code{return} at the end of function bodies
	\item only a basic set of operators; no bitwise or shift operators
	\item no function calls
	\item no cryptocurrency features (e.g., \code{transfer} and \code{payable})
\end{itemize}

\subsection{Usability}

\paragraph{Transaction Transformation}
Automatic transformation of transactions in \oldname is very limited and does
involve direct interaction with a blockchain. In particular, the public keys of
all involved parties and the current blockchain state need to be manually
provided to zkay together with the transaction parameters. Issuing the resulting
transformed transaction is not part of \oldname and needs to be done separately.
Overall, issuing a \oldname transaction is much more complicated than issuing a
standard Ethereum transaction.

\paragraph{Uninterpretable Compiler Errors}
In general, error reporting in \oldname is very limited and error messages often
only consist of internal stack traces not easily interpretable for users.

\paragraph{No Standardized Contract Distribution}
\oldname does not provide a specification how to distribute a zkay contract to
its users. Each user of a contract at least needs access to the original zkay
source code and all involved NIZK proving keys, which need to be distributed
using off-chain communication.

\subsection{Modularity and Extensibility}

The implementation of \oldname is not very modular, which impedes extensibility
of the tool. In particular, the compiler is tightly coupled to
ZoKrates~\cite{zokrates-github}, making it difficult to integrate other NIZK
frameworks. Additionally, key and cipher text sizes are hard-coded.

\subsection{Performance}

Off-chain computation and compilation performance of \oldname is suboptimal and
reduces development productivity. In particular, compiling a simple contract can
take up to several minutes.
\section{Architecture}

\begin{figure}
	\centering
	\includegraphics[width=1\textwidth]{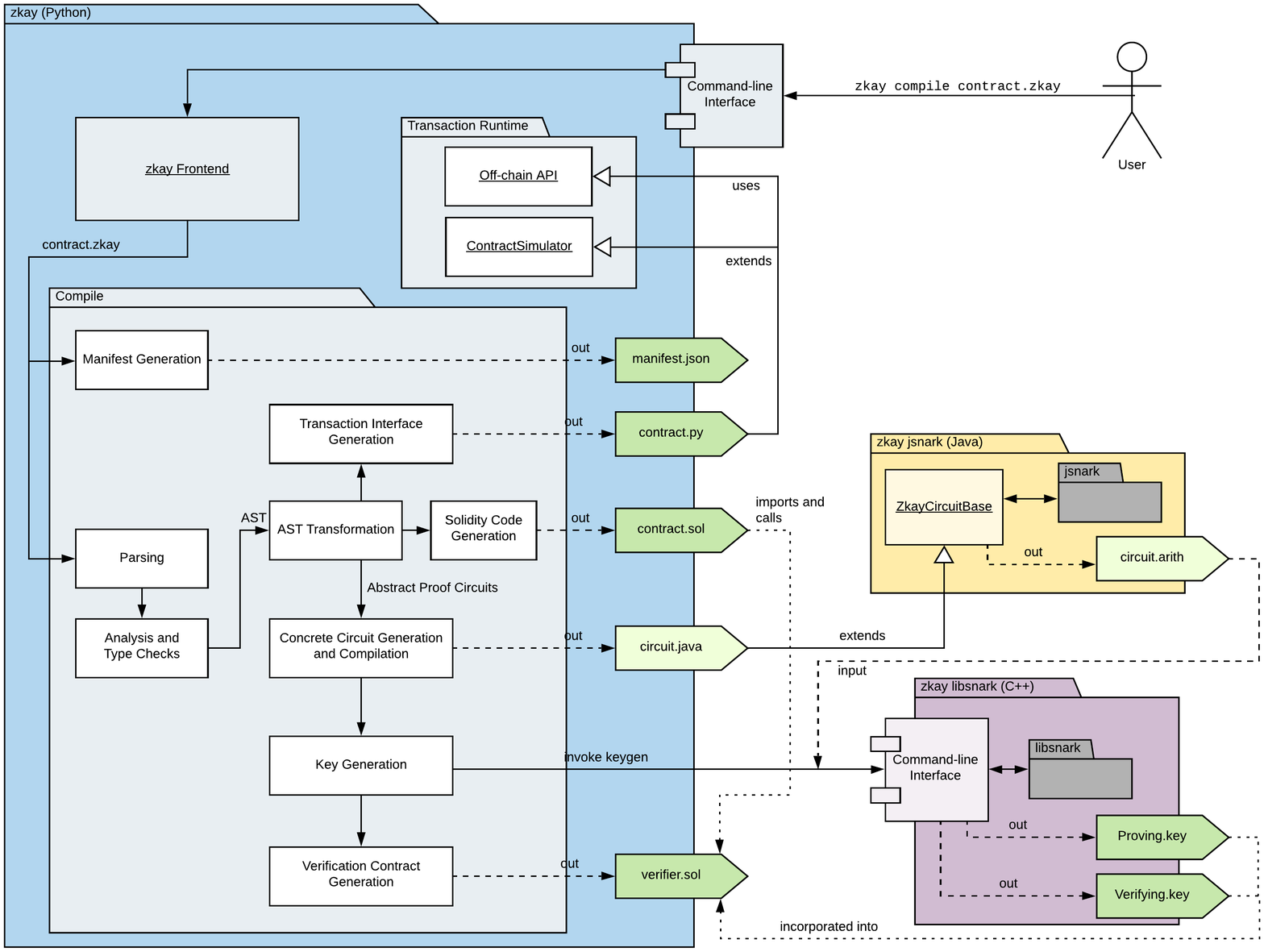}
	\caption{Architecture overview and compilation pipeline of \name.}
	\label{fig:zkay-compilation}
\end{figure}

This section describes the architecture of \name, which consists of a compiler
compiling zkay to Ethereum contracts (\cref{sec:arch-compilation}) and a runtime
for transaction transformation (\cref{sec:arch-transactions}).
\cref{fig:zkay-compilation} provides an overview of the architecture and the
interactions between the different components.

\subsection{Compiler}\label{sec:arch-compilation}

The compiler accepts a zkay contract and produces various output files as
described below (see dark green items in \cref{fig:zkay-compilation}).
\begin{itemize}
	\item \emph{contract.sol:} The main transformed Solidity contract performing
	on-chain computation (except proof verification, see next). This contract
	is deployed to the blockchain.
	\item \emph{verifier.sol:} Solidity contracts (one per proof circuit)
	performing NIZK proof verification on-chain, also deployed to the
	blockchain. The main contract calls these verification contracts.
	\item \emph{Proving.key, Verifying.key:} NIZK proof key pairs (one key pair
	per verification contract). The prover keys are used by users of the main
	contract to generate valid NIZK proofs. The verification keys are integrated
	into the verification contracts (see previous).
	\item \emph{manifest.json:} A manifest file storing all zkay compiler
	settings such that the compiler output can be exactly reproduced later.
\end{itemize}

As indicated in \cref{fig:zkay-compilation}, the compiler first parses the input
zkay contract, and performs various type checks and analyses to ensure the
contract's validity. Next, the compiler transforms the abstract syntax tree
(AST) of the contract to generate Solidity code and abstract proof circuits.
These circuits are then transformed to concrete circuits used as inputs for the
zk-SNARKs tools jsnark~\cite{jsnark-github} and libsnark~\cite{libsnark-github}.
After generating NIZK key pairs, the compiler generates the verification
contracts. From the transformed AST, the compiler also generates a transaction
interface that will later be used together with the transaction runtime.
Finally, the compiler also generates a manifest file.

\begin{figure}
	\centering
	\includegraphics[width=1\textwidth]{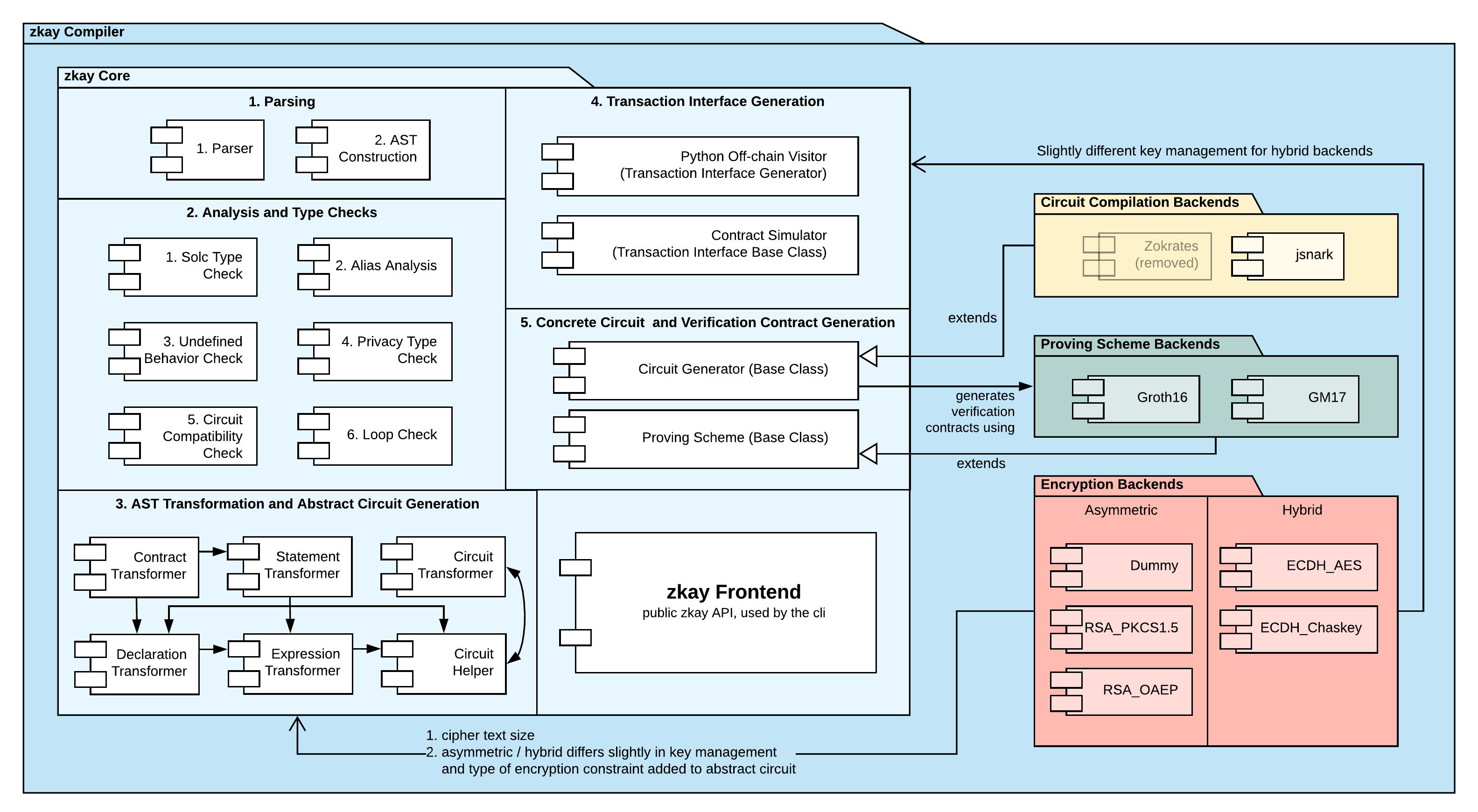}
	\caption{Architecture of the zkay compiler and its interaction with
	different backends.}
	\label{fig:zkay-compiler-arch}
\end{figure}

\cref{fig:zkay-compiler-arch} depicts the building blocks realising these
phases. The following sections discuss the compiler phases in more detail.

\subsubsection{Parsing, Analysis and Type Checks}\label{sec:parsing-analysis-type-checks}
The input contract is parsed using an extension of the grammar used in \oldname,
supporting the new language features described in \cref{sec:newlang}. Then, we
run several type checks and analyses on the constructed AST (see steps 1 and 2
in \cref{fig:zkay-compiler-arch}).

\paragraph{Solc Type Check} This check ensures that when ignoring all privacy
features specific to zkay, the contract is a valid Solidity contract. It (i)
replaces all comments and privacy features in the input contract by a matching
amount of whitespace, (ii) invokes the solc Solidity compiler via its json
interface, and (iii) displays any warnings or errors from solc in the context of
the original code. Due to the whitespace replacement, the source code locations
match the locations in the zkay contract.

Users can perform the ``stripping'' of zkay features in isolation using the
\code{zkay solify} command. This allows easy use of linters and other program
analysis tools designed for Solidity.

The remaining analyses only have to deal with zkay's privacy types and features.
For example, we don't need to re-implement Solidities data type checks.

\paragraph{Alias Analysis} Like in \oldname, the compiler performs alias
analysis on the contract to determine references which are guaranteed to point
to the caller's address at runtime. This information is later used by the privacy
type check. Compared to \oldname, the precision of the alias analysis in \name
is slightly more precise. For example, the join operation no longer removes
final address variables from equivalence sets.

\paragraph{Undefined Behavior Check} This check ensures that the zkay contract
is free of undefined behavior due to expressions relying on subexpression
evaluation order. In particular, the check forbids expressions where two
subexpressions have side-effects on the same variable, or where a variable is
written and read in two different subexpressions.

\paragraph{Privacy Type Check} Here, the compiler checks zkay's privacy types as
described in \cite{zkay}. For instance, it ensures that no implicit information
leaks are possible and that the zkay contract is realizable using encryption and
NIZK proofs. The typing rules of \oldname have been extended to account for the
new language features of \name (see \cref{sec:newlang}). Circuit compatibility
and loops are analyzed in two separate checks, see below.

\paragraph{Circuit Compatibility Check} This check ensures that private
expressions do not contain private operations that are not expressible inside a
NIZK proof circuit. Since \name supports function calls (see \cref{sec:fcalls}),
this involves (i) recursively checking the bodies of any called functions, and
(ii) ensuring that the body of any function called within a private expression
does not involve recursion or loops. Also, the compiler enforces private
expressions to be side-effect free in order to simplify circuit optimization.

\paragraph{Loop Check} Finally, the compiler checks that all loops in the
contract are fully public, meaning that no private expressions appear within any
loop exit condition or body.

\subsubsection{AST Transformation and Abstract Circuit Generation}\label{sec:ast-transform-ac-generation}
Next, the analyzed AST is transformed to yield abstract proof circuits and
Solidity code for the main contract as described below (see
\cref{fig:zkay-compilation} and step 3 in \cref{fig:zkay-compiler-arch}).

\paragraph{AST Transformation} \Name transforms the AST using an extension of
the translation rules from~\cite{zkay} to produce representations of the
on-chain computation and proof circuits.

\paragraph{Code Generation}
During code generation, zkay traverses the on-chain computation AST and emits
Solidity code. The resulting Solidity file is the main contract to be deployed
on the blockchain. Due to the improved modularity, additional target languages
can easily be incorporated in \name by adapting the code generator.

\paragraph{Abstract Circuit Generation}
Unlike \oldname, where contract transformation is heavily coupled with the
ZoKrates framework~\cite{zokrates,zokrates-github}, \name constructs
framework-agnostic abstract representations of proof circuits from the
transformed AST. Such an \emph{abstract proof circuit} maintains a list of
public and private circuit inputs, and holds a sequence of abstract circuit
statements from the following list.

\begin{itemize}
	\item \emph{Variable Declaration and Assignment:} Assigns the (plaintext) value of a
	private expression to a new temporary circuit variable.

	\item \emph{Guard Condition Modification:} Adds or removes a boolean circuit
	variable to resp.\ from the guard condition (see \S5.3 in~\cite{zkay} for
	more information on guard conditions).

	\item \emph{Encryption/Decryption Constraint:} An assertion of the form
	$cipher == \text{Enc}(plain, rnd, k)$, ensuring correct encryption of a
	plaintext $plain$ using key $k$ and randomness $rnd$. Such constraints are
	used for private function arguments and whenever the result of a private
	expression is stored to a private variable.
	
	An analogous decryption assertion exists, which is used whenever a private
	variable is read within a private expression.
	
	\item \emph{Equality Constraint:} An assertion of the form $val1 == val2$.
	Such constraints are used whenever a private value is declassified.
	
	\item \emph{Function Call:} A pointer to the abstract circuit of a called
	function. The concrete circuit generator (see below) will inline the target
	circuit at this position.
\end{itemize}

Note that there is no arbitrary assignment statement for proof circuits. \Name
relies on static single assignments, using the ``Variable Declaration and
Assignment'' statement. For simplicity, only this type of statement can contain
arbitrary expressions. All other statements may only reference variables
directly. Therefore, any compound expression is first assigned to a temporary
circuit variable before being used in any of the other statement types.

\subsubsection{Transaction Interface Generation}
The transformed AST is further processed to generate a transaction interface for
the contract. More specifically, this step generates Python code serving as a
user interface to the contract and transforming transactions by generating NIZK
proofs and encrypting function arguments.

\Cref{list:zk-oc-gen-example,list:zk-oc-gen-interface} show the structure of the
generated Python code for an example contract. At a high level, zkay creates a
Python class which matches the original contract structure, where each method
internally performs transaction transformation (transaction simulation, argument
encryption, proof generation) for the corresponding contract function. Users can
naturally interact with objects of this python class as if they would interact
with the contract directly. The generated class makes heavy use of transaction
runtime functionality provided by the superclass \code{ContractSimulator}. See
\cref{sec:contract-py} for more details.

\begin{figure}
	\begin{minipage}[b]{0.38\textwidth}
	\begin{minted}{\zkay}
	contract Test {
		constructor(uint x) {
			// ...
		}

		function f(uint@me val) {
			// ...
		}
	}






	\end{minted}
	\captionof{listing}{Example zkay contract.}
	\label{list:zk-oc-gen-example}
	\end{minipage}\hfill
	\begin{minipage}[b]{0.54\textwidth}
		\begin{minted}{python}
		class Test(ContractSimulator):
			@staticmethod
			def deploy(x: int, *, account: Addr) -> Test:
				# ...

			@staticmethod
			def connect(address: Addr, account: Addr) -> Test:
				# ...

			def f(val: int):
				val = enc(val, my_pk)
				zk_out = [..]
				# ...
				proof = gen_proof('f', zk_in, zk_out)
				transact('f', val, zk_out, proof)
		\end{minted}
		\captionof{listing}{Generated transaction interface (pseudocode).}
		\label{list:zk-oc-gen-interface}
	\end{minipage}
\end{figure}

This design has some important advantages over the design of \oldname.
\begin{itemize}
	\item \emph{Interpretability}: The generated Python code can be manually
	inspected by users to see how transactions are transformed.
	\item \emph{Simplified Debugging}: To debug contracts (and also the zkay
	implementation itself), one can use standard Python debugging tools.
	\item \emph{Improved Performance}: Since there is no additional overhead for
	interpreting zkay code (as done in \oldname), performance is slightly
	better.
\end{itemize}

\subsubsection{Concrete Circuit Generation and Compilation}\label{sec:circ-gen}
In this step, abstract proof circuits are compiled to a NIZK-framework-specific
concrete circuit representation. \Name uses different \emph{circuit compilation
backends} to target different NIZK frameworks. Each backend must implement
zkay's abstract \code{CircuitGenerator} interface, which provides functions for
(i)~transforming abstract circuits into a backend-specific representation with
equivalent semantics, (ii)~generating prover and verification keys for a given
circuit and proving scheme, and (iii)~marshalling generated verification keys
into a standard format only depending on the proving scheme (but not the
particular backend).

\Name currently supports only one circuit compilation backend, which relies on
the jsnark framework~\cite{jsnark-github} and is described below. The ZoKrates
integration from \oldname has been removed due to ZoKrates' lack of
cryptographic primitives.

\paragraph{Jsnark Backend} Jsnark is one of the few currently available
frameworks with comprehensive built-in support for cryptographic primitives. In
\name, a concrete jsnark circuit is represented as a Java subclass of
\code{CircuitGenerator}. It uses a Java API (see below) to (i)~define public and
private circuit input wires, and (ii)~combine wires using a variety of boolean,
arithmetic, and bitwise operations. The jsnark backend generates this Java class
and executes it to generate an output file (\code{circuit.arith} in
\cref{fig:zkay-compilation}) in libsnark-specific format. Then, a separate
libsnark~\cite{libsnark-github} interface binary (see below) is used to generate
prover and verification keys for the circuit.

\paragraph{Zkay Jsnark API} To increase readability and reduce the amount of
boilerplate code in the generated circuit files, \name implements a higher-level
Java API~\cite{zkay-jsnark-github} on top of jsnark. In particular, the API
provides the following items.
\begin{enumerate}
	\item A custom abstract \code{CircuitGenerator} subclass
	\code{ZkayCircuitBase}, which includes helper functions to define encryption
	constraints, add circuit inputs with a specific emulated type, and reference
	input wires by name.
	\item Adapter classes for jsnark's cryptographic primitive gadgets, which
	unpack/pack the inputs/outputs as required for zkay.
	\item A \texttt{TypedWire} wrapper class, which associates data types to
	wires and correctly emulates the semantics of arithmetic overflows, signed
	operations, etc.\ (see \cref{sec:integer-variants}).
\end{enumerate}

As an example, \Cref{list:zk-jsnark-api-ex} shows the jsnark Java circuit
corresponding to the zkay function in \cref{list:zk-jsnark-api-ex-zkay} and
making use of the jsnark Java API.

\begin{listing}
	\begin{minted}[]{\zkay}
	 function buy(uint amount) public {
	     require(registered[me]);
	     balance[me] = balance[me] + amount;
	 }
	\end{minted}
	\caption{Example zkay code.}
	\label{list:zk-jsnark-api-ex-zkay}
\end{listing}

\begin{listing}

	\begin{minted}[baselinestretch=1,fontsize=\tiny]{java}
	public class ZkayCircuit extends ZkayCircuitBase {
		public ZkayCircuit() {
			super("zk__Verify_Token_buy", "dummy", 248, 3, 1, 3, true);
		}

		private void __zk__buy() {
			stepIn("_zk__buy");
			addS("secret0_plain", 1, ZkUint(256));
			addS("zk__in0_cipher_R", 1, ZkUint(256));
			addS("zk__out0_cipher_R", 1, ZkUint(256));
			addIn("zk__in0_cipher", 1, ZkUint(256));
			addIn("zk__in1_plain_amount", 1, ZkUint(256));
			addOut("zk__out0_cipher", 1, ZkUint(256));

			//[ --- balance[me] + reveal(amount, me) ---
				// secret0_plain = dec(balance[me]) [zk__in0_cipher]
				checkDec("secret0_plain", "glob_key_me", "zk__in0_cipher_R", "zk__in0_cipher");
				// zk__in1_plain_amount = amount
				decl("tmp0_plain", o_(get("secret0_plain"), '+', get("zk__in1_plain_amount")));
				// zk__out0_cipher = enc(tmp0_plain, glob_key_me)
				checkEnc("tmp0_plain", "glob_key_me", "zk__out0_cipher_R", "zk__out0_cipher");
			//] --- balance[me] + reveal(amount, me) ---

			stepOut();
		}

		@Override
		protected void buildCircuit() {
			super.buildCircuit();
			addK("glob_key_me", 1);

			__zk__buy();
		}

		public static void main(String[] args) {
			ZkayCircuit circuit = new ZkayCircuit();
			circuit.run(args);
		}
	}
	\end{minted}
	\caption{Circuit using high-level jsnark Java API.}
	\label{list:zk-jsnark-api-ex}
\end{listing}

\paragraph{Zkay Libsnark Interface}
The original libsnark interface included in jsnark is not very flexible as it
e.g. does not support the GM17~\cite{gm17} proving scheme. For that reason,
\name uses its own custom C++ libsnark interface~\cite{zkay-libsnark-github},
which extends the existing interface with further proving schemes and
serialization of verification keys.

\subsubsection{Verification Contract Generation}\label{sec:verify-gen} In this
step, zkay generates Solidity contracts performing on-chain NIZK proof
verification. As the format of such verification contracts depends on the used
proving scheme, \name uses different \emph{proving scheme backends} for this
task. Each backend implements the abstract \code{ProvingScheme} interface and is
responsible for (i)~defining a verification key data structure, and
(ii)~generating verification contracts for a given verification key and list of
public circuit inputs. \Name provides backends for two proving schemes:
GM17~\cite{gm17}, which is already used by \oldname, and the more efficient
Groth16~\cite{groth16} scheme (default in \name).

In contrast to \oldname, where verification contracts are directly generated by
ZoKrates, the proving scheme backends in \name have full control over the
verification contracts and can apply specific optimizations such as loop
unrolling, avoiding unnecessary copy operations, and hashing optimizations (see
\cref{sec:hash-opt}).

\subsection{Transaction Transformation Runtime}\label{sec:arch-transactions}

The second core component of \name is a transaction runtime which transforms
function calls and executes them on the blockchain. It consists of two parts:
the Python contract interface generated by the compiler (see
\cref{sec:contract-py}) and the actual runtime providing core functionality via
an API (see \cref{sec:zkay-runtime}). In \cref{sec:interacting-with-contract},
we describe how users can interact with a zkay contract using its Python
interface, either programmatically or via an interactive shell.

\subsubsection{Python Contract Interface}\label{sec:contract-py} The Python
contract interface (\code{contract.py} in \cref{fig:zkay-compilation})
transforms and forwards function calls to the Solidity contract deployed on the
blockchain (\code{contract.sol}). Its goal is to prepare encrypted function
arguments and generate NIZK proofs to be accepted by the verifier contracts.
Because these generally depend on blockchain state and the results of public or
private operations in the zkay contract, the interface needs to simulate the
execution of the zkay contract in Python.

For many simple operations, the compiler simply emits the transformed AST as
equivalent Python code. However, the contract interface also performs more
complex operations as described below, often leveraging the zkay runtime API
(see \cref{sec:zkay-runtime}).

\begin{itemize}
	\item To make transaction transformation transparent to users, the signature
	of each external function matches the original zkay function. Its body
	encrypts private arguments and adds their plaintext values to the secret
	circuit arguments. At the beginning of the function, the \code{msg},
	\code{block} and \code{tx} objects are populated with current blockchain
	data using the runtime API.
	\item At the end of each external function that requires verification, the
	interface uses the runtime API to generate a NIZK proof for the collected
	circuit arguments. Then, the transformed transaction is issued using the
	runtime API.
	\item Parameter or function names conflicting with Python keywords or other
	reserved names are sanitized by adding a special suffix that is prevented to
	be used in user code.
	\item Key lookups in the PKI are replaced by a runtime API call
	retrieving the requested key from the blockchain state.
	\item The values of state variables are lazily retrieved from the blockchain
	and cached for repeated access. Each state variable reference is replaced by
	an index operation into a dedicated state dictionary that requests the value
	from the blockchain (via the runtime API) if it is not cached.
	\item Due to the missing support for nested local scopes in Python, \name
	uses context managers and a special local variable dictionary aware of
	scoping (provided by the runtime API) to emulate nested block scopes.
	\item Whenever a private circuit value is included in the circuit inputs, it
	is immediately decrypted (using the runtime API) and its plaintext value is
	added to the secret circuit arguments.
	\item For each private expression, the simulator computes the expression's
	output value, encrypts it (using the runtime API) if required, and stores it
	in the corresponding circuit output variable.
	\item Each require statement is replaced by an if-statement raising a
	\code{RequireException} if the condition does not hold.
	\item For all arithmetic operations and type casts, correct over-/underflow
	behavior is emulated using the runtime API (see \cref{sec:integer-variants}
	for details).
	\item Two additional static methods \code{connect} and \code{deploy} are
	added to the contract interface. The function \code{connect} verifies the
	integrity of the remote Solidity contract at the specified address (see
	\cref{sec:contract-integrity}) and creates a Python interface for it. The
	function \code{deploy} calls the contract constructor, which results in a
	deployment transaction. See \cref{sec:interacting-with-contract} for
	details.
\end{itemize}

\subsubsection{Zkay Runtime}\label{sec:zkay-runtime}

The interface \code{contract.py} extends the base class
\code{ContractSimulator}, which provides access to the zkay runtime API and
maintains the transaction simulation's internal state. The API provides core
functionality such as access to local and state variables, type casting and
under-/overflow emulation, blockchain interaction, cryptographic operations, and
key management (see \cref{sec:contract-py}).

\begin{figure}
	\centering
	\includegraphics[width=1\textwidth]{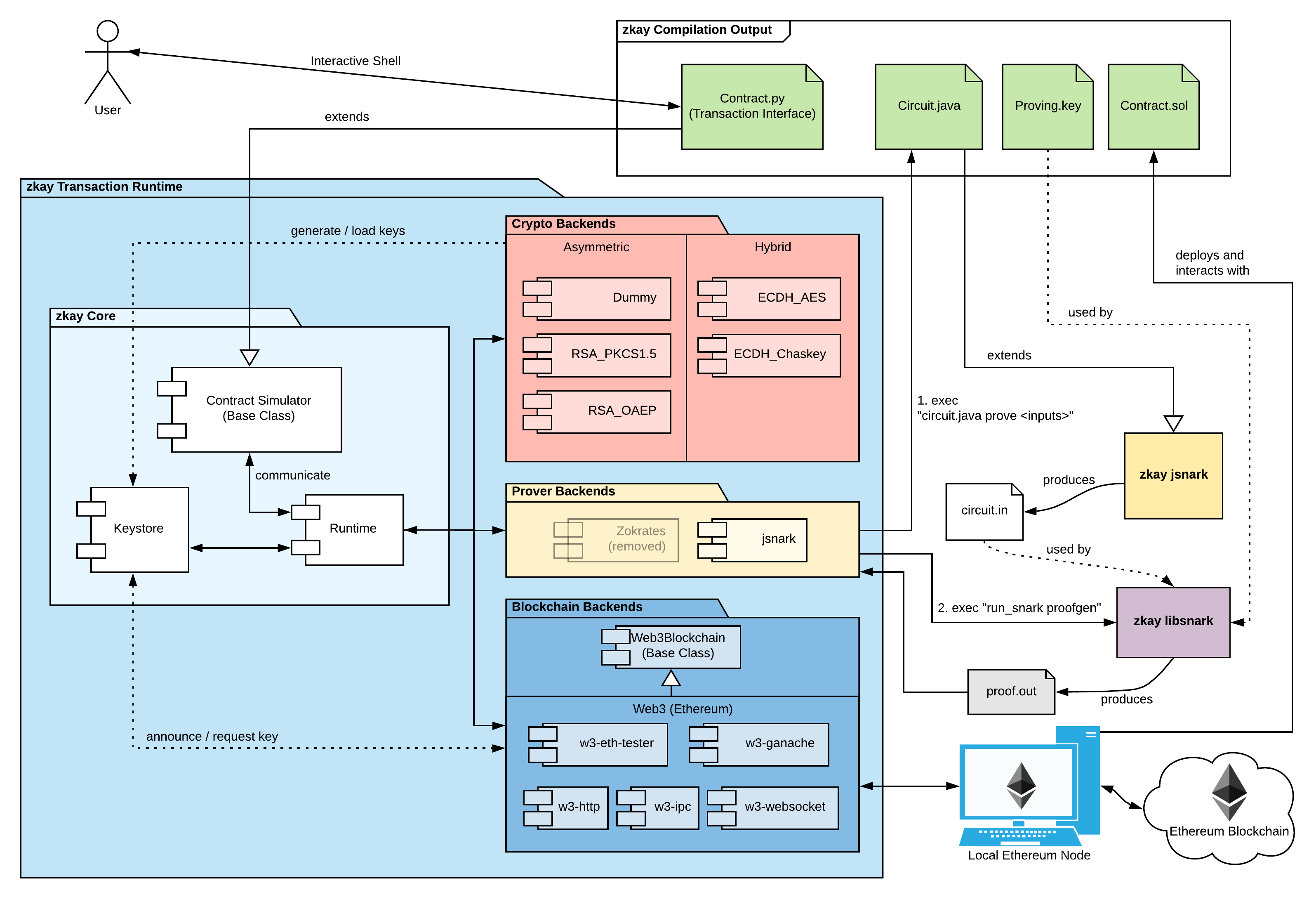}
	\caption{Architecture of the zkay transaction runtime.}
	\label{fig:zkay-transact-arch}
\end{figure}

\cref{fig:zkay-transact-arch} shows an overview of the transaction runtime
architecture. \Name supports different encryption schemes, which are realized
using different \emph{crypto backends}. To generate NIZK proofs, the runtime
further uses a \emph{prover backend}. The runtime can interact with a variety of
Ethereum blockchain interfaces, which are exposed to the runtime as
\emph{blockchain backends}. We next describe these backends.

\paragraph{Crypto backends}
These backends implement key generation, encryption and decryption operations
for different encryption schemes. \Name includes 5 crypto backends: ``dummy
encryption'' (the insecure surrogate encryption function used in \oldname), RSA
with either PKCS1.5~(\cite{pkcs1-rfc}, section~7.2) or OAEP~(\cite{pkcs1-rfc},
section~7.1), as well as ECDH~\cite{DH,ECDH,ECDH2} in combination with
AES~\cite{aes-standard} or Chaskey LTS~\cite{chaskey-cipher} block ciphers. See
\cref{sec:crypto} for details.

\paragraph{Prover backends}
A prover backend is responsible for generating NIZK proofs for a given circuit
and input values. \Name supports only one prover backend, which is based on
jsnark. To construct a NIZK proof, the jsnark backend first executes the Java
circuit (\code{circuit.java} in \cref{fig:zkay-transact-arch}) produced by the
jsnark circuit generator in proof generation mode to create a jsnark input file
(\code{circuit.in}). This input file is then passed to the libsnark interface,
which performs the actual proof generation.

\paragraph{Blockchain backends}
These backends connect zkay with different Ethereum blockchain interfaces. They
implement contract deployment, transactions, state queries (e.g., state variable
and account balance queries), and integrity checks (see
\cref{sec:contract-integrity}). All five currently supported blockchain backends
are based on web3py~\cite{web3py-github}. The
\texttt{w3-eth-tester}~\cite{ethtester} and \texttt{w3-ganache}~\cite{ganache}
backends are used for testing with local blockchains, while the remaining
backends are used to connect zkay with real Ethereum nodes over IPC, HTTP, or
WebSocket. For security reasons, zkay prevents using dummy encryption for
non-local blockchains.

\subsubsection{Using the Contract Interface}\label{sec:interacting-with-contract}

Users can use the \code{deploy} and \code{connect} commands of zkay's
command-line interface to easily create and interact with \name contracts. Using
\code{deploy}, a user can deploy a zkay contract to the configured Ethereum
blockchain. Then, an instance of the contract interface can be obtained using
the \code{connect} command. When running this command, zkay enters an
interactive shell in the context of the created interface object, which allows
users to issue transactions in an interactive manner. All external contract
functions are available with identical signature as in the zkay contract. When
calling any such function, \name transparently transforms the transaction (see
\cref{sec:contract-py}) and issues it on the configured Ethereum blockchain.

The \code{deploy} and \code{connect} functions can also be accessed using the
programmatic interface in the module \code{zkay.zkay\_frontend}.

\section{New Language Features}\label{sec:newlang}

In this section, we describe the new language features introduced in
\name.

\subsection{Function Calls}\label{sec:fcalls}

As a major extension of \oldname, \name supports internal function calls.

\subsubsection{Restrictions}\label{sec:calls-restrictions}

Expressions with side-effects (e.g., an expression modifying a state variable)
cannot be moved to the proof circuit. Inside private expressions (i.e., if there
exists a non-public ancestor in the expression tree), \name hence enforces
called functions to be annotated as \code{pure} or \code{view}.

In general, calls to functions with private return value are inlined in the
proof circuit if they occur within a private expression. As a result, the called
function bodies may not contain any operations unsupported in the proof circuit
(such as loops or recursive function calls). This is enforced by the type system
of \name. In contrast, calls to functions with public return value are generally
not integrated in the proof circuit. If such a function call appears inside a
private expression, the return value is computed on-chain and passed to the
proof circuit as a public input.

\subsubsection{Function Calls not Requiring
Verification}\label{sec:calls-no-verif}

There are two cases where an internal function call does not require any
verification, namely if (i)~the function is fully public, or (ii)~the only
private expressions in the function are its private arguments.~\footnote{Case
(ii) particularly applies to a function that merely stores the value of its
private argument in a state variable. While calls to such functions require
verification when being called externally (the correct encryption of private
user-provided arguments is checked in the proof circuit), this is not the case
for internal calls.} There are no restrictions on such function calls, as they
are not transformed by \name in any way.

\subsubsection{Function Calls Requiring Verification}\label{sec:calls-with-verif}

Internal function calls outside private expressions require verification if the
called function contains private expressions beyond any private arguments. In
general, there are two ways to ensure that the required proof circuit is
included in a NIZK proof and verified:

\begin{enumerate}
	\item\label{opt-callee} \emph{Callee-driven}. Here, the callee is
	responsible for verifying the proof circuit induced by its own body,
	excluding nested function calls. With each nested function call, an
	additional NIZK proof is introduced that has to be tunneled through the
	caller.
	\item\label{opt-caller} \emph{Caller-driven}. Here, the caller is
	responsible for verifying the proof circuit induced by \emph{any}
	transitively called function. In this case, the external top-level function
	performs verification of a single large proof circuit combined from smaller
	sub-circuits induced by nested function calls. Only a single NIZK proof is
	introduced per external function.
\end{enumerate}

As the gas cost for verifying a NIZK proof on the blockchain is virtually
independent of the proof circuit size (see \cref{sec:hash-opt}), option
\ref{opt-caller}, which induces less proof verifications, is more efficient.
Accordingly, \name follows the caller-driven approach. For this approach, the
following items have to be considered.

\begin{itemize}
	\item For each external function, the compiler needs to statically determine
	all transitively called internal functions and construct the required proof
	circuit. For this to work, the type system has to guarantee the absence of
	recursive calls and function calls within loops (see
	\cref{sec:parsing-analysis-type-checks}).

	\item External functions need access to the public proof circuit inputs and
	outputs for all transitively called internal functions.

	\item Each internally called function needs access to its own circuit
	outputs.
	
	\item Verifying correct encryption of private function parameters is
	required if and only if the function is called externally. In all other
	cases (internal or private functions), the function can rely on the fact
	that its arguments have already been verified by the caller.
\end{itemize}

\paragraph{Circuit Input and Output Arrays}
In order to manage public circuit inputs, \name uses a large dynamic array
shared by transitively called functions. This array is (implicitly) divided into
a hierarchy of sections according to the tree of nested function calls. Each
section stores the function's own circuit inputs as well as the sections of all
called functions. An analogous second array is used for circuit outputs.

\begin{figure}
	\centering
	\includegraphics[width=0.8\textwidth]{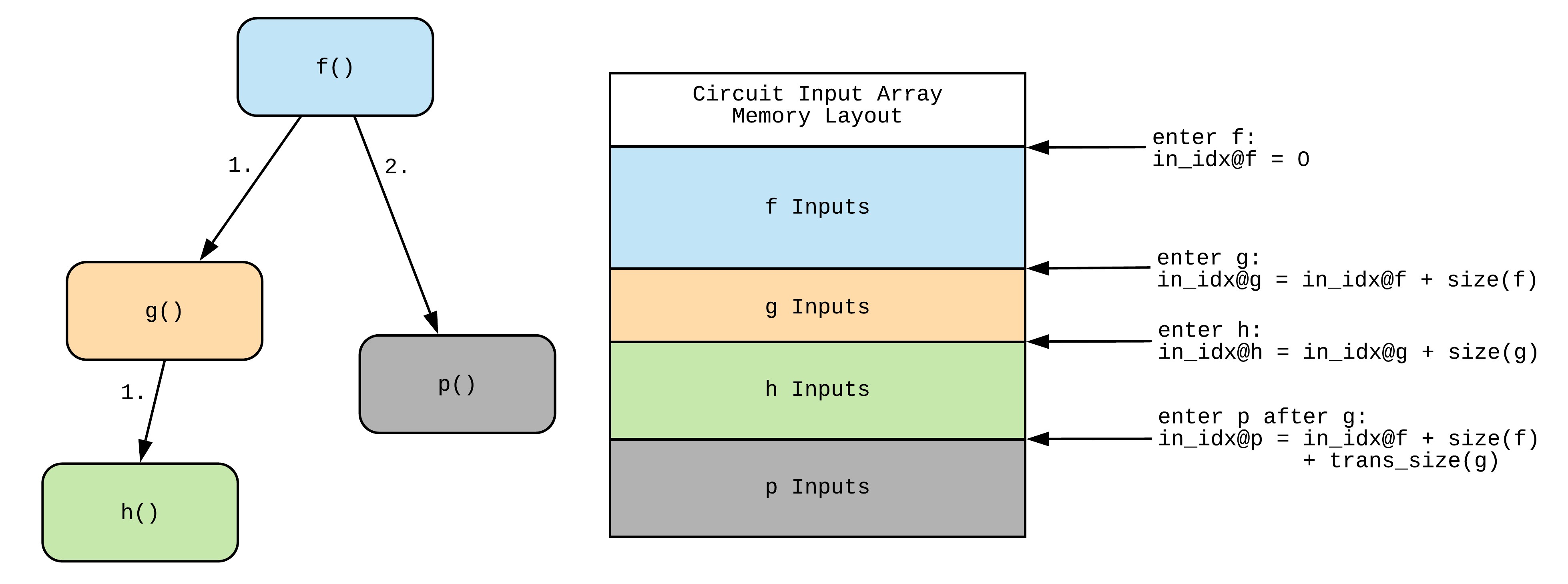}
	\caption{Memory layout of proof circuit input array.}
	\label{fig:zkay-call-memlayout}
\end{figure}

\cref{fig:zkay-call-memlayout} visualizes the input array memory layout for an
example call tree. Here, function \code{f} is called externally and calls
functions \code{g} and \code{p} in its function body.

\paragraph{Functions with Private or Internal Visibility Modifier} 
These functions cannot be called externally. \Name adds four parameters to the
signature of such functions: dynamic arrays \code{in} and \code{out} used to
propagate circuit inputs (resp.\ outputs) along the call tree, and two integer
indices \code{in\_idx} and \code{out\_idx} used to indicate the section start
offsets for the current function.

The indices \code{in\_idx} and \code{out\_idx} act as relocation base addresses
and are set by the caller of the function. The transformed Solidity code
accesses entries of the \code{in} and \code{out} array relative to these
offsets, which makes it possible to use the same function definition
independently of the function's location in the call tree.

To work around Solidity's stack size limit and lack of array slice operators,
the circuit outputs are deserialized from the \code{out} array into a struct at
the beginning of the function body. Analogously, the public circuit inputs are
serialized from a struct into the \code{in} array before the function returns.
Within the function body, all circuit outputs resp.\ inputs are read from resp.\
stored into the corresponding struct.

\paragraph{Functions with Public Visibility Modifier} Functions with public
visibility modifier may be called both internally and externally. Hence, these
are split into two functions during compilation.

\begin{itemize}
	\item An internal function, which is simply a copy of the original function
	transformed as any other internal function (see above).
	
	\item An external function with two additional parameters \code{proof} and
	\code{out}, allowing the user to pass a NIZK proof and a circuit output
	array according to the memory layout described above. The body of the
	external function performs the following steps.

	\begin{enumerate}[topsep=4pt,leftmargin=20pt]
		\item Allocate an array large enough to store the circuit inputs of all
		transitively called functions.

		\item Request all encryption keys required by any transitively called
		function from the PKI contract.

		\item Store all encrypted parameters in the circuit input array (the
		correct encryption of these parameters will be verified).

		\item Call the corresponding internal function (see above), passing the
		\code{in} and \code{out} arrays with initial indices. The called
		function will populate the \code{in} array.

		\item Invoke the NIZK proof verifier for \code{proof}, \code{in} and
		\code{out}.
	\end{enumerate} 
\end{itemize}

\paragraph{Incorporating Circuits of Nested Calls}
The zkay compiler generates a separate abstract proof circuit (see
\cref{sec:ast-transform-ac-generation}) for each function definition. Abstract
proof circuits use a special \code{CircuitCall} statement to include
(sub-)circuits of nested function calls. The circuit compilation backend (see
\cref{sec:circ-gen}) is then responsible for inlining the sub-circuits into the
main top-level proof circuit, respecting the memory layout of the \code{in} and
\code{out} arrays.

\begin{figure}
	\centering
	\includegraphics[width=1.0\textwidth]{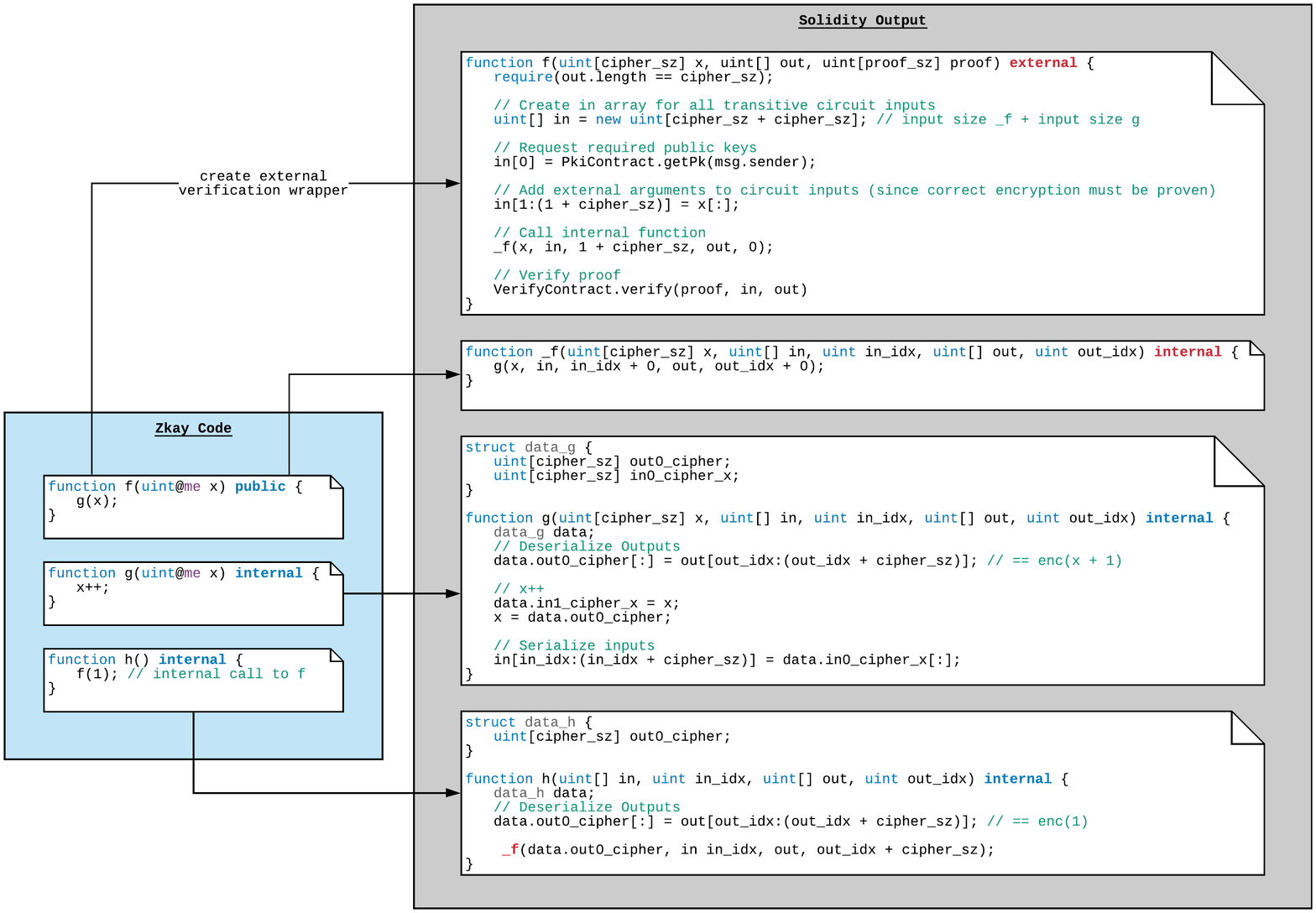}
	\caption{Simplified example of compiling nested function calls to Solidity.}
	\label{fig:zkay-fcall-trafo}
\end{figure}

\paragraph{Example} \Cref{fig:zkay-fcall-trafo} demonstrates how \name complies
function calls to Solidity. To remove clutter, range checks are omitted and a
Python-style slicing syntax is used to indicate array ranges.

\subsection{Cryptocurrency Functionality}\label{sec:cc-features}

\Name allows functions to be declared \code{payable}. Further, it supports
address types (see \cref{sec:address-and-enum-types}) and querying the balance
of an address using the member function \code{balance}. Outside private
expressions, payable addresses can be used to transfer funds using the member
functions \code{send} and \code{transfer}.

Further, \name provides access to the \code{msg} (only \code{sender} and
\code{value}), \code{block}, and \code{tx} globals. During transaction
simulation (see \cref{sec:arch-transactions}), they are represented as Python
objects, which are populated by the blockchain backend.

\subsection{If-Statements}\label{sec:if-statement}

\Name supports control-flow using if-statements, both with public and private
conditions.

\subsubsection{Public Conditions}
\Name adds support for if-statements with public conditions and makes use of
guard conditions as introduced in~\cite{zkay}: if any of the branches contains
private expressions, an appropriate guard condition is added to all proof
circuit constraints generated for that branch.

For example, consider the zkay code in \cref{lst:if-condition}. Its transformed
version and generated proof circuit are shown in
\cref{lst:if-condition-transformed}. Using the implication \code{guard =>
(enc(3) == out[0])} for the assertion ensures that the condition \code{enc(3) ==
out[0]} is only checked in the proof circuit if the guard condition \code{p ==
2} is true.

\begin{figure}
\begin{minipage}[b]{0.3\textwidth}
	\begin{minted}{\zkay}
		uint p = ...;
		uint@me o;
		if (p == 2) {
			o = 3;
		}
		o = o + 1;


	\end{minted}
	\captionof{listing}{If-condition in zkay.}\label{lst:if-condition}
\end{minipage}\hfill
\begin{minipage}[b]{0.63\textwidth}
	\begin{minted}{\zkay}
	uint p = ...;
	uint[cipher_sz] o;  // proof circuit:
	in[0] = (p == 2);   //  guard = in[0]
	if (in[0]) {
	  o = out[0];       //  assert guard => (enc(3) == out[0])
	}
	in[1:] = o;         //  tmp = in[0] + 1
	o = out[1];         //  assert enc(tmp) == out[1] 
	\end{minted}
	\captionof{listing}{Transformed code (simplified).}\label{lst:if-condition-transformed}
\end{minipage}
\end{figure}

\subsubsection{Private Conditions}
If-statements with private conditions are supported, however with some
restrictions making sure that the visible trace of a transaction does not leak
any information about the value of the condition. In particular, no branch is
allowed to contain operations unsupported in proof circuits or any side-effects
apart from assignments to primitive type variables which are private to the
caller.

Zkay first collects the set $X$ of all variables which are assigned in at least
one branch. Then, it replaces the if-condition by an assignment to all variables
in $X$. The correctness of the right-hand side is verified in the proof circuit,
which (i) evaluates both branches of the if-statement, and (ii) uses conditional
assignment expressions to select the appropriate values for the variables in $X$
according to the value of the condition. In particular, all variables in $X$ get
assigned new re-encrypted values, even if the underlying plaintext values were
not modified in the branch taken.

\subsection{Short-circuit Evaluation}

Certain operators such as \code{||} and \code{\&\&} are subject to short-circuit
evaluation in Solidity (i.e., not all operands are necessarily evaluated). \Name
correctly handles this using appropriate guard conditions in the proof circuits.

For example, consider the zkay code in \cref{lst:shortcirc-req-guard-eg}, where
the function call \code{priv()} is skipped if \code{b} is true due to
short-circuit evaluation. When creating the proof circuit for \code{test}, zkay
adds a guard condition to the encryption assertion in \code{priv} to make sure
it is only checked if \code{b} is false.

\begin{listing}
	\begin{minted}{\zkay}
	function priv() returns(bool) {
		uint@me v = 2;   // encryption assertion is only checked in the proof circuit if !b
		return true;
	}

	function test(bool b)  {
		bool val = b || priv();
	}
	\end{minted}
	\caption{Guard condition necessary due to short-circuit evaluation.}
	\label{lst:shortcirc-req-guard-eg}
\end{listing}

\subsection{Public Loops}

\Name supports \code{while}, \code{do} ... \code{while} and \code{for} loops
whose condition, update statement and body do not contain any private
expressions or calls to functions requiring verification (see
\cref{sec:fcalls}).

\subsection{Tuples}

\Name supports Solidity tuples. This can e.g. be used to pack multiple return
values or swap variables. Nesting and mixing values with different privacy types
is supported. As a tuple is only a syntactic group, it does not have a privacy
type itself.

\subsection{Integer Variants}\label{sec:integer-variants}

\Name provides advanced support for integer variants, including fixed-sized (see
\cref{sec:integer-emu}) and signed (see \cref{sec:integer-signed}) integers.

The zk-SNARKs frameworks underlying the circuit compilation backends require all
proof circuit operations to be expressed using finite arithmetic in a prime
field. Prime field numbers are inherently unsigned and restricted in size by the
field prime. For instance, the field prime involved in the elliptic
Barreto-Naehrig curve ``alt\_bn128'' used in
libsnark~\footnote{\url{https://github.com/scipr-lab/libff/tree/master/libff/algebra/curves/alt_bn128}
(accessed 2020-08-25).} is roughly 253.5 bits in size, making it impossible to
accurately represent 256-bit operations. Directly translating Solidity
\code{uint} operations to operations in the prime field (as done in \oldname) is
hence incorrect.

\Name correctly emulates the semantics of different integer variants (e.g.,
signed arithmetic, under- and overflow behavior) in the used prime field whenever possible. Otherwise, compiler errors or warnings are raised.

\subsubsection{Integer Sizes}\label{sec:integer-emu}

Like Solidity,  \name supports different integer sizes (i.e., \code{uint8},
\code{uint16}, ..., \code{uint248}, \code{uint256}).

\paragraph{Up to 248 bits}
For sizes up to 248 bits, the jsnark circuit compilation backend emulates
correct overflow behavior using the low-level finite field operation primitives
provided by jsnark.

\begin{itemize}
	\item \emph{Addition} Emulated by restricting the field addition output
	(which can have at most 249 bits and hence does not overflow at the field
	prime) to the desired bit width.

	\item \emph{Negation}: Emulated by constructing the ``two's complement'',
	see \cref{sec:integer-signed}.

	\item \emph{Subtraction}: Emulated by adding the negated value.

	\item \emph{Multiplication}: The result of multiplying two $n$-bit integers
	can have up to $2n$ bits, which means that a field prime overflow can
	already occur when multiplying integers with as little as $n \geq 128$ bits.
	For this reason, multiplication of ($n\geq128$)-integers is emulated using
	multiple $\frac{n}{2}$-bit multiplications, whose results do not overflow.
\end{itemize}

\paragraph{256 bits}
\Name uses a single prime field element to represent 256-bit integers, even
though this leads to a semantic mismatch between prime field and Solidity
operations. In particular:

\begin{enumerate}
	\item all private arithmetic operations overflow at the field prime, and
	\item private comparison operations fail for values $\geq2^{252}$.
\end{enumerate}

Because private 256-bit operations are safe as long as they only involve values
below $2^{252}$, \name does not generally forbid such operations. Instead, it
raises an according compiler warning to make developers aware of potential
issues.

\subsubsection{Signed Integers}\label{sec:integer-signed}

\Name uses the ``two's complement'' representation for signed integers and
correctly emulates signed integer arithmetic for at most $248$ bits. Private
256-bit signed integers are not supported.

\subsection{Address and Enum Types} \label{sec:address-and-enum-types}

\Name supports \code{address} and \code{address payable} variables, both of
which can be private. Cryptocurrency-related members such as \code{balance}
(see \cref{sec:cc-features}) are only accessible on public addresses.

Further, \name supports declaring and using custom \code{enum} types. These are
fully supported inside private expressions.

\subsection{Type Casts}

Like Solidity, \name allows explicit type conversions between many primitive
types. Also, \name reflects the implicit type conversions of Solidity (e.g.,
conversions from smaller to larger integer sizes). Type casts are valid
operations inside private expressions. The privacy type of a type cast
expression is inherited from its source expression.

\subsection{More Operators}

\paragraph{Assignment Operators}
\Name introduces assignment operators such as \code{+=} and \code{*=}. These are
syntactic sugar and can only appear as statements (i.e., they are not
expressions).~\footnote{Because of Solidity's unspecified expression evaluation
order, treating assignment operators as expressions would lead to unspecified
behavior and side-effects.} One notable exception are loop update expressions,
where assignment operators can be used for convenience.

\paragraph{Pre- and Post-increments}

\Name supports \code{++} and \code{--} as prefix or postfix operators. Similarly
as assignment operators, pre- and post-increments are statements and cannot be
used as expressions, except for loop update expressions.

\paragraph{Bitwise Operators}
The bitwise operators \code{\&}, \code{|}, \code{\textasciitilde} and
\code{\textasciicircum} are supported by \name, except for private 256-bit
integers.

\paragraph{Shifts}
The shift operators \code{<<} and \code{>>} are supported by \name, except for
private 256-bit integers. For public operands, there are no further
restrictions. When shifting private values, the shift amount must be a public
constant.

\section{Security Features}\label{sec:crypto}

In this section, we present the security features of \name. In particular, we
show how zkay employs asymmetric (\cref{sec:enc-asymmetric}) and
hybrid (\cref{sec:enc-hybrid}) encryption to protect private data, and how
private variables are initialized in this context (\cref{sec:enc-default-init}).
Further, we describe how zkay checks the integrity of deployed contracts
(\cref{sec:contract-integrity}).

\subsection{Asymmetric Encryption}\label{sec:enc-asymmetric}

\Name supports asymmetric encryption, where data is encrypted under the owner's
public key.

\subsubsection{Public Key Infrastructure}\label{sec:asymmetric-pki} Each user
account $a$ creates an asymmetric key pair $(\text{sk}_a, \text{pk}_a)$, whose
public part $\text{pk}_a$ is published to a dedicated zkay public key
infrastructure (PKI) contract. Whenever a compiled zkay contract requires the
public key of an address, it requests the key from the PKI contract.

\subsubsection{Verifying Encryption}
Often, a compiled zkay contract needs to verify that some plaintext $m$ was
correctly encrypted for some target address $a$. More specifically, for a given
ciphertext $c$, the contract must ensure that $c$ was obtained by encrypting the
plaintext $m$ with $a$'s public key $\text{pk}_a$ and some secret
randomness~$r$.

\cref{fig:example-compile-crypto-asymmetric} exemplifies how \name verifies
correctness of encryption for an asymmetric encryption backend. The code in
\cref{lst:example-compile-crypto-asymmetric-input} is compiled to the Solidity
code in \cref{lst:example-compile-crypto-asymmetric-sol}, which verifies that
\code{c} is the result of encrypting \code{val} under the public key of
\code{other} and some secret randomness \code{rnd}. This is captured in the
proof circuit shown in \cref{lst:example-compile-crypto-asymmetric-circuit},
which takes public circuit inputs \code{m} and \code{pk}, the private circuit
input \code{rnd}, and returns the (public) output \code{c}.

\begin{figure}
	\centering
	\begin{minipage}[b]{0.4\textwidth}
		\begin{minted}{\zkay}
		function f(uint val) {
			uint@other x = val;
		}




		\end{minted}
		\captionof{listing}{Example zkay code.}\label{lst:example-compile-crypto-asymmetric-input}
	\end{minipage}\hspace{1cm}
	\begin{minipage}[b]{0.40\textwidth}
		\begin{minted}{\zkay}
		function f(uint val, Cipher c, proof) {
			uint m = val;
			Key pk = PKI.get(other);
			Cipher x = c; // c == enc(val)

			verify([m, pk], [c], proof);
		}
		\end{minted}
		\captionof{listing}{Transformed Solidity code.}\label{lst:example-compile-crypto-asymmetric-sol}
	\end{minipage}\\\vspace{0.3cm}
	\begin{minipage}[b]{0.40\textwidth}
		\begin{minted}{python}
		def f(val):
			m = val
			pk = request_key(other)
			c, rnd = enc(m, pk)
			x = c

			proof = prove_f([rnd], [m, pk], [c])
			transact('f', [val, c, proof])
		\end{minted}
		\captionof{listing}{Python contract interface.}
	\end{minipage}\hspace{1cm}
	\begin{minipage}[b]{0.40\textwidth}
		\begin{minted}{yaml}
		priv_in: rnd
		pub_in: m, pk
		pub_out: c

		proof:
		 - assert c == enc(m, pk, rnd)


		\end{minted}
		\captionof{listing}{Proof circuit (pseudocode).}\label{lst:example-compile-crypto-asymmetric-circuit}
	\end{minipage}
	\caption{Verifying correct encryption for an asymmetric encryption backend.}\label{fig:example-compile-crypto-asymmetric}
\end{figure}

\subsubsection{Verifying Decryption}\label{sec:enc-asymmetric-verify-dec}
Whenever a private variable \code{c} is used in a private expression $e$, the
value of \code{c} must be decrypted to its plaintext value $m$ by the contract
interface to allow local computation of the (plaintext) value of~$e$. The
correctness of this decryption operation is checked in the proof circuit.

For many asymmetric encryption schemes, decryption is more expensive than
encryption. Hence, \name checks correctness using the inverse \emph{encryption}
operation in the proof circuit. More specifically, it checks whether the value
of \code{c} can be obtained by encrypting the plaintext $m$ with the owner's
public key and some secret randomness. This is exemplified in
\cref{fig:example-compile-crypto-asymmetric-decrypt}.

\begin{figure}
	\begin{minipage}[b]{0.40\textwidth}
		\begin{minted}{\zkay}
		function f() {
			uint@me val;
			uint x = reveal(val, all);
		}





		\end{minted}
		\captionof{listing}{Example zkay code.}
	\end{minipage}\hspace{1cm}
	\begin{minipage}[b]{0.40\textwidth}
		\begin{minted}{\zkay}
		function f(uint m, Proof proof) {
			Cipher val;

			Cipher c = val;
			Key pk = PKI.get(msg.sender);
			uint x = m;

			verify([c, pk], [m], proof);
		}
		\end{minted}
		\captionof{listing}{Transformed Solidity code.}
	\end{minipage}\\\vspace{0.3cm}
	\begin{minipage}[b]{0.40\textwidth}
			\begin{minted}{python}
			def f():
				val = Cipher()

				c = val
				pk = request_key(msg.sender)
				m, rnd = dec(val, get_sk(msg.sender))
				x = m

				proof = prove_f([rnd], [c, pk], [m])
				transact('f', [m, proof])
			\end{minted}
			\captionof{listing}{Python contract interface.}
		\end{minipage}\hspace{1cm}
		\begin{minipage}[b]{0.40\textwidth}
			\begin{minted}{yaml}
			priv_in: rnd
			pub_in: c, pk
			pub_out: m

			proof:
			- assert c == enc(m, pk, rnd)




			\end{minted}
			\captionof{listing}{Proof circuit (pseudocode).}
		\end{minipage}
		\caption{Verifying correct decryption for an asymmetric encryption backend.}\label{fig:example-compile-crypto-asymmetric-decrypt}
\end{figure}

\subsubsection{Available Backends}
\Name supports two asymmetric crypto backends described below, both of which are
based on RSA~\cite{rsa}. The backends rely on
PyCryptodome's~\cite{pycryptodome-github} RSA implementation for the contract
interfaces, and jsnark's RSA gadges for the proof circuits.

\paragraph{RSA PKCS1.5}

This backend uses RSA encryption with PKCS\#1 v1.5 padding
(see~\cite{pkcs1-rfc}, section~7.2). In comparison to OAEP padding (see below),
PKCS1.5 padding offers better off-chain performance (see
\cref{sec:perf-benchmarks}).

\paragraph{RSA OAEP}
This backend uses RSA encryption with PKCS\#1 v2.0 OAEP padding
(see~\cite{pkcs1-rfc}, section~7.1).

\subsubsection{Limitations}
RSA crypto backends come with some efficiency drawbacks (see
\cref{sec:perf-benchmarks} for an experimental evaluation). In particular:
\begin{itemize}
	\item RSA requires large key and cipher text sizes. In particular, these
	items have at least 2048 bits, which amounts to 9 proof circuit inputs. As
	the cost for NIZK proof verification is linear in the number of public proof
	circuit inputs, this leads to high gas costs.

	\item RSA encryption, particularly with OAEP, is a very expensive operation
	in the proof circuit. This leads to high memory consumption and long
	execution times for proof generation.
\end{itemize}
For these reasons, \name also offers hybrid encryption as described in the next
section.

\subsection{Hybrid Encryption}\label{sec:enc-hybrid}

\Name supports hybrid encryption, where an elliptic curve (EC) key exchange is used
in combination with a symmetric block cipher. At a high level, hybrid encryption
in \name works as follows.

\begin{itemize}
	\item Each account $a$ creates an asymmetric EC key pair $(\text{sk}_a,
	\text{pk}_a)$, whose public part $\text{pk}_a$ is published in the PKI.
	\item Whenever an account $a$ wants to encrypt data for a target account
	$b$, it uses elliptic curve Diffie-Hellman (ECDH)~\cite{DH,ECDH,ECDH2} to
	obtain a shared secret from $\text{sk}_a$ and $\text{pk}_b$. From this
	shared secret, a shared symmetric key $k_{a,b} = k_{b,a}$ is derived.
	\item Account $a$ then uses a symmetric block cipher to encrypt the
	plaintext using key $k_{a,b}$ and obtain the ciphertext $c$, which includes
	an initialization vector (IV). The ciphertext is extended by the public key
	of $a$ to obtain the tuple $(c, \text{pk}_a)$, which is stored on the
	blockchain.
	\item When decrypting a ciphertext tuple $(c, \text{pk}_a)$, account $b$
	uses ECDH to obtain $k_{a,b}$ from $\text{sk}_b$ and the public key
	$\text{pk}_a$ in the tuple. Then, $b$ can decrypt $c$ using $k_{a,b}$.
\end{itemize}

\subsubsection{Public Key Infrastructure}
The PKI for hybrid encryption works analogously as for asymmetric encryption
(see \cref{sec:asymmetric-pki}). EC keys are much smaller than RSA keys for
equivalent levels of security. In particular, for the elliptic curve supported by
jsnark, a strong EC key equivalent to a 2048-bit RSA key fits within a single
proof circuit input.

\subsubsection{Verifying Encryption}
When verifying that a ciphertext tuple $(c, \text{pk}_a)$ is the result of
encrypting a plaintext $m$ for a target address $b$, the compiled zkay contract
must ensure that (i)~$c$ was obtained by encrypting $m$ with the key $k_{a,b}$
derived from $\text{pk}_b$ and some secret key $\text{sk}_a$, and
(ii)~$\text{pk}_a$ and $\text{sk}_a$ form an EC key pair.

\cref{fig:example-compile-crypto-hybrid} exemplifies how \name verifies
correctness of encryption for a hybrid encryption backend. The Solidity code in
\cref{lst:example-compile-crypto-hybrid-sol} verifies that \code{c} is the
result of encrypting \code{val} for the account \code{other}. In particular, it
sets the public key for \code{c} to the public key of the
sender,~\footnote{Technically, the implementation uses a \code{uint[3]} array to
combine the IV, actual cipher text $c$ and public key $\text{pk}_a$ of the
originator $a$ instead of a ciphertext tuple $(c, \text{pk}_a)$.} loads the
public key of \code{other}, and calls the verifier. The proof circuit shown in
\cref{lst:example-compile-crypto-hybrid-circuit} takes as public arguments the
public key \code{pk\_me} of the sender, the plaintext \code{m}, and the public
key \code{pk} of \code{other}. As a private argument, it takes the secret key
\code{sk\_me} of the sender. The circuit (i)~asserts that \code{pk\_me} and
\code{sk\_me} form an EC key pair, (ii)~constructs the symmetric key \code{k}
shared between the sender and \code{other} using ECDH, and (iii)~asserts
\code{c} has been encrypted correctly.

\begin{figure}
	\begin{minipage}[b]{0.40\textwidth}
		\begin{minted}{\zkay}
		function f(uint val) {
			uint@other x = val;
		}







		\end{minted}
		\captionof{listing}{Example zkay code.}
	\end{minipage}\hspace{1cm}
	\begin{minipage}[b]{0.40\textwidth}
		\begin{minted}{\zkay}
		function f(uint val, IvCipher c, proof) {
			Key pk_me = PKI.get(msg.sender);

			c.src = pk_me;
			uint m = val;
			Key pk = PKI.get(other);
			IvCipher x = c; // c == enc(val)

			verify([pk_me, m, pk], [c], proof);
		}
		\end{minted}
		\captionof{listing}{Transformed Solidity code.}\label{lst:example-compile-crypto-hybrid-sol}
	\end{minipage}\\\vspace{0.3cm}
	\begin{minipage}[b]{0.40\textwidth}
			\begin{minted}{python}
			def f(val):
				pk_me = request_key(msg.sender)
				sk_me = get_sk(msg.sender)

				m = val
				pk = request_key(other)
				iv_c = enc(m, ecdh(pk, sk_me))
				x = iv_c

				proof = prove_f(
					[sk_me], [pk_me, m, pk], [iv_c]
				)
				transact('f', [val, iv_c, proof])

			\end{minted}
			\captionof{listing}{Python contract interface.}
		\end{minipage}\hspace{1cm}
		\begin{minipage}[b]{0.40\textwidth}
			\begin{minted}[escapeinside=||,mathescape=true]{yaml}
			priv_in: sk_me
			pub_in: pk_me, m, pk
			pub_out: iv_c

			proof:
			 - # once per circuit
			 - assert pk_me == G |$\cdot$| sk_me

			 - # once per pk per circuit
			 - k = ECDH(sk_me, pk)

			 - # for each encryption
			 - iv = iv_c[:128] # first 128 bits
				assert iv_c == [iv, enc(m, k, iv)]
			\end{minted}
			\captionof{listing}{Proof circuit (pseudocode).}\label{lst:example-compile-crypto-hybrid-circuit}
		\end{minipage}
		\caption{Verifying correct encryption for a hybrid encryption backend.}\label{fig:example-compile-crypto-hybrid}
\end{figure}

\subsubsection{Verifying Decryption}
Like for asymmetric encryption (see \cref{sec:enc-asymmetric-verify-dec}), zkay
proves correct decryption using an encryption operation in the proof circuit.
Verifying decryption works similarly as for encryption (see
\cref{fig:example-compile-crypto-hybrid}), however the shared encryption key is
derived from the originator's public key stored as part of the ciphertext.

\subsubsection{Available Backends}
\Name supports two hybrid crypto backends described below. The shared key is
derived from the ECDH shared secret by taking the 128 leftmost bits of its
SHA-256 digest~\cite{sha2-standard}. As an optimization, shared keys are only
computed once per proof circuit.

\paragraph{ECDH AES}
This backend combines SHA-256-ECDH key derivation with
AES-128~\cite{aes-standard} encryption in CBC mode~\cite{cbc-patent}. The
implementation relies on PyCryptodome's AES-CBC implementation and jsnark's CBC,
AES and ECDH gadgets.

\paragraph{ECDH Chaskey}
This backend is a more lightweight alternative to ECDH AES. It uses the Chaskey
LTS block cipher~\cite{chaskey-cipher}, which is natively supported by jsnark
and allows for more efficient proof circuits than AES. The implementation relies
on jsnark's CBC, Chaskey and ECDH gadgets. The contract interface combines
BouncyCastle's~\cite{bouncycastle} CBC mode with a custom Chaskey LTS block
cipher implementation.

\subsection{Default Initialization for Private Variables}\label{sec:enc-default-init}

In Solidity, variables are implicitly initialized with zero-equivalent values
(e.g., reading an uninitialized integer variable is guaranteed to return zero).
While this is a useful feature often leveraged by Solidity contracts, it is an
issue for compiled zkay contracts, where private variables would have to be
implicity initialized with the \emph{encryption of zero}. For this reason,
\oldname restricts support for default initialization and has undefined behavior
when reading uninitialized private variables.

\Name adds support for zero-initialized private variables. The general idea is
to rely on Solidity's initialization but treat private variables with values
zero as if they were encrypted. More specifically, proof circuit assertions of
the form  $cipher = enc(plain, k)$ are actually implemented as $(cipher = 0
\Rightarrow plain = 0) \wedge (cipher \neq 0 \Rightarrow cipher = enc(plain,
k))$. Also, decrypting the ciphertext~$0$ in the contract interface is
configured to return the plaintext $0$. Finally, an additional constraint
asserting that user-provided ciphertexts are never $0$ is included in proof
circuits. In the extremely unlikely event that encrypting a plaintext in the
contract interface leads to the ciphertext~$0$, the plaintext is re-encrypted
using fresh randomness.

\subsection{Checking Contract Integrity}\label{sec:contract-integrity}

In order to interact with a compiled zkay contract deployed on the blockchain, a
user needs access to the following items: (i)~the original zkay contract,
(ii)~the NIZK prover and verifier keys, and (iii)~the manifest file specifying
the originally used compiler version and options. These items provide enough
information to locally reconstruct the contract interface. \Name comes with
utilities to bundle these items in an archive that can be distributed to and
imported by users via an out-of-band channel.

\paragraph{Integrity of Remote Contracts}
When using a deployed contract, users need to verify that it corresponds
to the local archive they have obtained via an out-of-band channel. This is, the
remote EVM bytecode must be verified to match the result of compiling (i) using
(ii--iii).

\Name automatically performs this verification whenever the Python contract
interface is attached to a remote contract $C$ by the \code{connect} command
(see \cref{sec:interacting-with-contract}). In particular, it performs the
following steps.

\begin{enumerate}
	\item Zkay compiles the local zkay contract using the compiler settings in
	the manifest file and a special mode that uses the existing prover and
	verifier keys instead of re-generating them. The compiler outputs the
	transformed Solidity contract $C_\text{local}$, all verification contracts,
	the PKI contract, and any required library contracts.

	\item The addresses of the remote PKI, library and verification contracts as
	used by the remote contract are retrieved from bytecode.

	\item The PKI and verifier address placeholders in $C_\text{local}$ are
	replaced with the concrete addresses of the corresponding remote contracts
	obtained in the previous step. Similarly, the remote library addresses are
	linked to the local verification contracts.
	
	\item $C_\text{local}$ and all locally generated verification contracts, PKI
	contract, and library contracts are compiled with solc using the compiler
	settings in the manifest file.

	\item The bytecode of $C$, the remote PKI, library and verification
	contracts is compared to the compilation results from the previous step. If
	the bytecode is not equal, zkay raises an error and aborts the connection.
\end{enumerate}

\paragraph{Trusted Setup Phase}
Like in any NIKZ-proof framework based on zk-SNARKs, prover and verifier key
generation in \name relies on a trusted setup phase. The entity running key
generation must destroy a secret string generated during this phase, commonly
referred to as ``toxic waste''. If this string is retained by a malicious user,
the user can construct arbitrary fake proofs accepted by the verifier and hence
break zkay's correctness guarantees (however, not it's privacy
guarantees)~\cite{fake-proofs}. The user creating and distributing a zkay
contract archive must therefore be trusted to execute the setup phase correctly
and destroy the toxic waste.

In practice, trusted setup phases are often implemented using complex ceremonies
based on secure multi-party computation in order to weaken trust
assumptions~\cite{zcash-paramgen}. While zkay users can manually establish such
ceremonies, \name does not come with built-in support for establishing suchlike.

\section{Performance}

Next, we present the various optimizations employed in \name to reduce
compilation time, off-chain runtime and memory requirements, and on-chain gas
costs (\cref{sec:perf-optimizations}). Further, we compare the performance of
\name to \oldname using a set of benchmarks (\cref{sec:perf-benchmarks}).

\subsection{Optimizations}\label{sec:perf-optimizations}

This section presents important optimizations employed by \name.

\subsubsection{Optimized Input Hashing}\label{sec:hash-opt}

The size of the verification key and the gas cost of NIZK proof verification are
linear in the number of public circuit inputs and outputs. The number of public
circuit inputs can be reduced by (i)~making public circuit inputs $\text{in}_1,
\ldots \text{in}_n$ private, and (ii)~providing a hash $h$ over these inputs as
a single public circuit input, which is computed on-chain and verified in the
proof circuit. This leads to low verification costs that are almost constant in
the number of circuit inputs, except for a small linearly-increasing cost for
computing $h$ on-chain.~\footnote{We note that this comes at the cost of
increased proof generation runtime and memory consumption.}

In \oldname, $h$ was computed as $h =
\text{sha256}(\text{sha256}(\text{sha256}(\ldots, \text{in}_{n-2}),
\text{in}_{n-1}), \text{in}_n)$, which results in $2n$ SHA-256 compressions for
$n$ public 256-bit circuit inputs.~\footnote{Each call of sha256 needs to hash a
512 bit payload (input and previous digest) plus another 512-bits due to
Merkle-Damgård length padding.} \Name improves this by using a single SHA-256
hash over the concatenated inputs: $h = \text{sha256}(\text{in}_1, \text{in}_2,
\ldots)$. This construction only requires $\lfloor\frac{n}{2}\rfloor + 1$
SHA-256 compressions for $n$ public 256-bit circuit inputs. Further, \name
provides a configurable threshold on the number $n$ of public circuit inputs
above which this construction should be applied.

\subsubsection{Constant Folding}

Like in Solidity, \name uses dedicated number literal types for constants and
applies constant folding. In particular, the value of a public constant
sub-expression within a private expression is computed at compile time and
integrated into the proof circuit.

\subsubsection{Circuit Input Caching}\label{sec:circ-input-cache}

In \oldname, multiple accesses of the same variable within a proof circuit led
to multiple redundant circuit inputs. \Name performs caching of circuit inputs
and re-uses these inputs for all accesses of the same variable provided it is
not modified. If a variable is modified in the zkay contract, the cache is
evicted and the variable is re-imported into the proof circuit.

\subsubsection{Public Key Caching}

\Oldname requests the required public key from the PKI and imports it using a
new circuit input every time it constructs an encryption or decryption
constraint in the proof circuit. If the same public key is required multiple
times, this leads to redundant circuit inputs and PKI lookups.

In \name, public keys are cached and only imported once whenever possible. In
particular, as the owner of a private variable remains constant during an entire
transaction, \name imports all public keys required for a transaction
\emph{once}, in the external top-level function.~\footnote{For tagged mapping
entries (e.g. \code{mapping (address!x => uint@x)}), this optimization is not
applied as the owner depends on the mapping index, which may change dynamically
at runtime.}

\subsubsection{Prover and Verifier Key Caching}

\Name caches generated prover and verifier keys to re-use them during
compilation if the corresponding proof circuit did not change since the previous
compilation. As key generation amounts for a large part of the compilation time,
this often leads to significantly lower compilation times and higher development
productivity.

\subsubsection{Parallelization}
\Name compiles different proof circuits in parallel to speed up compilation.
Also, it leverages the multi-threading support of libsnark for key and proof
generation.

\subsection{Benchmarks}\label{sec:perf-benchmarks}

\begin{table}
	\caption{Contract compilation time [s].}
	\label{tab:comp-runtime}

	\centering
	\begin{tabular}{l|rr|rrrrl@{}}
		\toprule
		& \multicolumn{2}{c|}{``dummy encryption''}& \multicolumn{5}{c}{\name}\\
		Contract         & \multicolumn{1}{l}{\oldname} & \multicolumn{1}{l|}{\name} & \multicolumn{1}{l}{ecdh-chaskey} & \multicolumn{1}{l}{ecdh-aes} & \multicolumn{1}{l}{rsa-pkcs1.5} & \multicolumn{1}{l}{rsa-oaep} &  \\ \midrule
		exam             & 291.91                        & 18.63                     & 61.77                            & 88.42                        & 318.17                          & 488.14                       &  \\
		income           & 124.27                        & 14.30                     & 41.23                            & 55.50                        & 162.82                          & 245.37                       &  \\
		insurance        & 416.16                        & 35.65                     & 104.74                           & 140.07                       & 467.49                          & 710.10                       &  \\
		lottery          & 83.66                         & 10.07                     & 34.28                            & 45.80                        & 96.33                           & 153.99                       &  \\
		med-stats        & 235.46                        & 16.35                     & 51.68                            & 70.88                        & 236.99                          & 356.09                       &  \\
		power-grid       & 123.05                        & 14.31                     & 44.75                            & 59.49                        & 192.59                          & 288.25                       &  \\
		receipts         & 183.11                        & 19.73                     & 63.54                            & 83.30                        & 263.37                          & 405.42                       &  \\
		reviews          & 305.95                        & 23.50                     & 78.35                            & 107.02                       & 357.54                          & 545.39                       &  \\
		sum-ring         & 123.17                        & 12.43                     & 45.69                            & 60.15                        & 179.96                          & 274.94                       &  \\
		token            & 295.18                        & 23.50                     & 72.82                            & 98.67                        & 331.52                          & 508.55                       &  \\ \midrule
		Mean             & 218.19                        & 18.85                     & 59.88                            & 80.93                        & 260.68                          & 397.62                       &  \\
		Speedup   & -                          & 11.58                     & 3.64                             & 2.70                         & 0.84                            & 0.55                         &  \\ \bottomrule
	\end{tabular}
\end{table}

In this section, we compare the performance of \name to its predecessor
\oldname. In particular, we compare compilation time, memory and output size
(\cref{sec:eval-compilation}); on-chain gas costs (\cref{sec:eval-on-chain});
and off-chain transaction runtime and memory (\cref{sec:eval-off-chain}). In
\cref{sec:eval-proving-schemes}, we compare the performance for different
proving schemes.

For our comparison in \crefrange{sec:eval-compilation}{sec:eval-off-chain}, we
use the GM17~\cite{gm17} proving scheme available in both implementations. We
evaluate \name on all its crypto backends (see \cref{sec:crypto}). This includes
``dummy encryption'', which enables a direct comparison to \oldname. We evaluate
both implementations on the 10 example contracts analyzed in zkay's original
publication~\cite[Tab.~1]{zkay}.

\clearpage
All experiments are conducted on a system with the following specifications.
\begin{itemize}
	\item \emph{CPU}: Intel i7-8700K 6x4.7GHz (+SMT)
	\item \emph{RAM}: 32 GB DDR4-3200
	\item \emph{OS}: MX Linux 19.1 x64, Kernel 4.19
	\item \emph{Compiler}: GCC 8.3.0, OpenJDK 11.0.7, Python 3.7.3
\end{itemize}

\subsubsection{Compilation Performance}\label{sec:eval-compilation}

We now analyze zkay's compilation time and memory requirements, as well as the
compilation output size.

\paragraph{Compilation Time}
\Cref{tab:comp-runtime} compares the compilation time of \oldname and \name with
different crypto backends for the 10 evaluated contracts. The majority of
compilation time is due to circuit compilation and key generation in the proving
scheme backends. As a result, the compilation times for different crypto
backends (which induce different proof circuit complexities) vary significantly.

\Name reduces the compilation time for ``dummy encryption'' by a factor of
around 11.6. Even though RSA backends induce much more complex proof circuits,
compilation times for these backends in \name are similar than for ``dummy
encryption'' in \oldname. This is due to the strong optimizations performed in
\name. ECDH based hybrid backends are more efficient and even allow for faster
compilation than \oldname with ``dummy encryption''.

\paragraph{Compilation RAM Usage}

\Cref{tab:comp-peak-mem} compares the peak RAM usage during compilation of the
10 evaluated contracts in \oldname and \name with different crypto backends. For
the insurance contract, the RSA PKCS1.5 backend required almost 25~GB (resp.\
10~GB) of RAM during circuit compilation (resp.\ key generation), which makes
compilation on low-end machines impossible. In contrast, hybrid crypto backends
are much more memory efficient.

\begin{table}
	\caption{Peak memory usage during compilation {[}MB{]}.}
	\label{tab:comp-peak-mem}

	\centering
	\begin{tabular}{l|rr|rrrrl@{}}
		\toprule
		& \multicolumn{2}{c|}{``dummy encryption''}& \multicolumn{5}{c}{\name}\\
		Contract         & \multicolumn{1}{l}{\oldname} & \multicolumn{1}{l|}{\name} & \multicolumn{1}{l}{ecdh-chaskey} & \multicolumn{1}{l}{ecdh-aes} & \multicolumn{1}{l}{rsa-pkcs1.5} & \multicolumn{1}{l}{rsa-oaep} &  \\ \midrule
		exam                    & 3299.56                       & 774.21                    & 2782.38                          & 6865.15                      & 19700.89                        & 19219.52                     &  \\
		income                  & 1639.13                       & 592.68                    & 1564.26                          & 7070.88                      & 8737.90                         & 11608.73                     &  \\
		insurance               & 3131.96                       & 1166.66                   & 2713.01                          & 11092.28                     & 24591.12                        & 23332.04                     &  \\
		lottery                 & 586.27                        & 672.40                    & 1116.64                          & 8201.20                      & 4427.00                         & 5501.43                      &  \\
		med-stats               & 3234.79                       & 580.10                    & 1899.68                          & 6233.57                      & 11719.60                        & 14476.78                     &  \\
		power-grid              & 1505.89                       & 768.29                    & 1574.00                          & 7275.74                      & 8329.46                         & 11176.63                     &  \\
		receipts                & 1411.74                       & 739.35                    & 1519.86                          & 9460.97                      & 9652.18                         & 16056.25                     &  \\
		reviews                 & 3820.18                       & 874.72                    & 3100.03                          & 5668.42                      & 18955.75                        & 22539.82                     &  \\
		sum-ring                & 1473.79                       & 511.28                    & 1831.28                          & 6844.18                      & 8752.44                         & 11597.67                     &  \\
		token                   & 3230.38                       & 946.04                    & 2342.00                          & 10268.23                     & 12664.50                        & 17110.29                     &  \\ \midrule
		Mean                    & 2333.37                       & 762.57                    & 2044.31                          & 7898.06                      & 12753.08                        & 15261.92                     &  \\
		Reduction Factor & -                          & 3.06                      & 1.14                             & 0.30                         & 0.18                            & 0.15                         &  \\ \bottomrule
	\end{tabular}
\end{table}

\begin{table}
	\caption{Compiled contract storage requirement {[}MB{]}.}
	\label{tab:storage-req}

	\centering
	\begin{tabular}{l|rr|rrrrl@{}}
		\toprule
		& \multicolumn{2}{c|}{``dummy encryption''}& \multicolumn{5}{c}{\name}\\
		Contract         & \multicolumn{1}{l}{\oldname} & \multicolumn{1}{l|}{\name} & \multicolumn{1}{l}{ecdh-chaskey} & \multicolumn{1}{l}{ecdh-aes} & \multicolumn{1}{l}{rsa-pkcs1.5} & \multicolumn{1}{l}{rsa-oaep} &  \\ \midrule
		exam                     & 1447                          & 286                       & 1299                             & 1754                         & 7245                            & 10940                        &  \\
		income                   & 581                           & 213                       & 795                              & 1004                         & 3594                            & 5276                         &  \\
		insurance                & 2027                          & 553                       & 2138                             & 2753                         & 10604                           & 15755                        &  \\
		lottery                  & 368                           & 137                       & 639                              & 776                          & 2146                            & 3266                         &  \\
		med-stats                & 1157                          & 252                       & 1057                             & 1370                         & 5246                            & 7790                         &  \\
		power-grid               & 581                           & 213                       & 881                              & 1118                         & 4231                            & 6177                         &  \\
		receipts                 & 870                           & 299                       & 1271                             & 1605                         & 5959                            & 8789                         &  \\
		reviews                  & 1519                          & 369                       & 1649                             & 2117                         & 8198                            & 12204                        &  \\
		sum-ring                 & 581                           & 186                       & 928                              & 1160                         & 3964                            & 5912                         &  \\
		token                    & 1443                          & 357                       & 1482                             & 1923                         & 7526                            & 11172                        &  \\ \midrule
		Mean                     & 1057.40                       & 286.50                    & 1213.90                          & 1558.00                      & 5871.30                         & 8728.10                      &  \\
		Reduction Factor & -                          & 3.69                      & 0.87                             & 0.68                         & 0.18                            & 0.12                         &  \\ \bottomrule
	\end{tabular}
\end{table}

\paragraph{Output Size}
\Cref{tab:storage-req} compares the storage of the compilation output. This is
dominated by the prover keys, whose size depends on the complexity of the proof
circuit. With RSA backends, prover keys become very large (several GB per
circuit), which makes contract distribution expensive. The compilation output of
hybrid backends is much smaller, but still in the order of 1~GB.

\paragraph{Summary}
In \cref{fig:comp-relations}, we visualize the relative differences in
compilation time, memory usage, and output size for the different backends
in \name. All values are normalized with respect to \oldname (which uses ``dummy
encryption''). In general, hybrid crypto backends are more efficient than RSA
backends for all considered metrics.

\begin{figure}
	\centering
	\begin{tikzpicture}
	\begin{axis}[
	ybar,
	width=0.9\textwidth,
	height=0.6\textwidth,
	ymax=900,
	ymin=0,
	minor tick num=3,
	ymajorgrids=true,
	yminorgrids=true,
	symbolic x coords={\oldname,dummy,ecdh-chaskey,ecdh-aes,rsa-pkcs1.5,rsa-oaep},
	ylabel={Normalized Value [\%]},
	legend style={at={(0.5,-0.15)}, anchor=north, legend columns=-1},
	]

	\addplot coordinates {(\oldname, 100.00) (dummy, 8.64) (ecdh-chaskey, 27.45) (ecdh-aes, 37.09) (rsa-pkcs1.5, 119.47) (rsa-oaep, 182.24)};
	\addplot coordinates {(\oldname, 100.00) (dummy, 32.68) (ecdh-chaskey, 87.61) (ecdh-aes, 338.48) (rsa-pkcs1.5, 546.55) (rsa-oaep, 654.07)};
	\addplot coordinates {(\oldname, 100.00) (dummy, 27.09) (ecdh-chaskey, 114.80) (ecdh-aes, 147.34) (rsa-pkcs1.5, 555.26) (rsa-oaep, 825.43)};
	\legend{Compilation Time, Peak Memory Usage,Output Storage Requirements}
	\end{axis}
	\end{tikzpicture}
	\caption{Mean compilation time, memory usage and output size for different backends in \name, normalized w.r.t.\ \oldname.}
	\label{fig:comp-relations}
\end{figure}

\subsubsection{On-chain Gas Costs}\label{sec:eval-on-chain}

We now analyze the on-chain costs of zkay transactions for a set of scenarios,
which comprise multiple transactions executed on the same contract (see
\cite{zkay} for details). \Cref{tab:trans-gas} compares the average transaction
gas costs for the scenarios using \oldname and \name with different crypto
backends. The numbers exclude the (one-time) costs for deployments and public
key announcements in the PKI contract. In \cref{fig:gas-efficiency}, we show the
mean transaction gas cost per crypto backend.

The ``dummy encryption'' backend in \name is around 11.7\% more gas-efficient
than \oldname. Also, the hybrid crypto backends result in lower transaction
costs. Only RSA backends lead to increased costs, likely due to the
significantly larger ciphertext and key sizes.

\begin{table}
	\caption{Average transaction gas cost (w/o deployment transactions) {[}gas{]}.}
	\label{tab:trans-gas}

	\centering
	\begin{tabular}{l|rr|rrrrl@{}}
		\toprule
		& \multicolumn{2}{c|}{``dummy encryption''}& \multicolumn{5}{c}{\name}\\
		Scenario         & \multicolumn{1}{l}{\oldname} & \multicolumn{1}{l|}{\name} & \multicolumn{1}{l}{ecdh-chaskey} & \multicolumn{1}{l}{ecdh-aes} & \multicolumn{1}{l}{rsa-pkcs1.5} & \multicolumn{1}{l}{rsa-oaep} &  \\ \midrule
		exam                  & 975867                        & 867507                    & 942841                           & 942916                       & 1368591                         & 1368527                      &  \\
		income                & 958761                        & 844096                    & 862339                           & 862323                       & 938187                          & 938043                       &  \\
		insurance             & 863896                        & 759457                    & 805235                           & 805219                       & 961517                          & 961525                       &  \\
		lottery               & 961293                        & 870737                    & 915360                           & 915440                       & 1313058                         & 1313090                      &  \\
		med-stats             & 963358                        & 841806                    & 874977                           & 874897                       & 996628                          & 996612                       &  \\
		power-grid            & 955134                        & 842065                    & 875727                           & 875684                       & 998738                          & 998674                       &  \\
		receipts              & 956674                        & 843245                    & 878341                           & 878266                       & 1002490                         & 1002565                      &  \\
		reviews               & 968579                        & 852586                    & 901632                           & 901632                       & 1071598                         & 1071521                      &  \\
		sum-ring              & 958754                        & 846932                    & 877017                           & 876985                       & 983056                          & 982944                       &  \\
		token                 & 971990                        & 855217                    & 890301                           & 890301                       & 1025587                         & 1025459                      &  \\ \midrule
		Mean                  & 953430.56                     & 842364.74                 & 882376.90                        & 882366.24                    & 1065945.00                      & 1065895.99                   &  \\
		Reduction Factor & -                          & 1.13                      & 1.08                             & 1.08                         & 0.89                            & 0.89                         &  \\ \bottomrule
	\end{tabular}
\end{table}

\begin{figure}
	\centering
	\begin{tikzpicture}
	\begin{axis}[
	ybar,
	width=0.9\textwidth,
	height=0.5\textwidth,
	ymin=500000,ymax=1200000,
	ymajorgrids,
	symbolic x coords={\oldname,dummy,ecdh-chaskey,ecdh-aes,rsa-pkcs1.5,rsa-oaep},
	ylabel={Mean Transaction Cost [gas]},
	legend style={at={(0.5,-0.15)}, anchor=north, legend columns=-1},
	]

	\addplot coordinates {(\oldname, 953431) (dummy, 842365) (ecdh-chaskey, 882377) (ecdh-aes, 882366) (rsa-pkcs1.5, 1065945) (rsa-oaep, 1065896)};
	\end{axis}
	\end{tikzpicture}
	\caption{Mean transaction gas cost (w/o deployment transactions) for different crypto backends.}
	\label{fig:gas-efficiency}
\end{figure}

\subsubsection{Off-chain Transaction Performance}\label{sec:eval-off-chain}

We now analyze zkay's off-chain performance when creating and issuing
transactions. More precisely, we analyze the runtime and peak memory usage of
issuing transactions using the contract interface of \name (resp.\ the
transaction transformation of \oldname). Both runtime and memory are dominated
by NIZK proof generation.

\paragraph{Runtime}
\Cref{tab:trans-runtime} compares the total runtime for creating and issuing all
transactions in the given scenarios using \oldname and \name with different
crypto primitives. The numbers include deployment transactions, NIZK proof
generation, and transaction execution on a local test blockchain (ganache-cli
for \oldname, eth-tester with py-evm backend for \name). \Name is faster than
\oldname for ``dummy encryption'' as well as for hybrid crypto backends.

\begin{table}
	\caption{Total off-chain runtime for executing all scenario transactions
	{[}s{]}.}
	\label{tab:trans-runtime}

	\centering
	\begin{tabular}{l|rr|rrrrl@{}}
		\toprule
		& \multicolumn{2}{c|}{``dummy encryption''}& \multicolumn{5}{c}{\name}\\
		Scenario         & \multicolumn{1}{l}{\oldname} & \multicolumn{1}{l|}{\name} & \multicolumn{1}{l}{ecdh-chaskey} & \multicolumn{1}{l}{ecdh-aes} & \multicolumn{1}{l}{rsa-pkcs1.5} & \multicolumn{1}{l}{rsa-oaep} &  \\ \midrule
		exam             & 243.47                        & 28.48                     & 102.13                           & 158.64                       & 512.09                          & 708.31                       &  \\
		income           & 115.14                        & 18.44                     & 52.44                            & 81.69                        & 208.82                          & 278.49                       &  \\
		insurance        & 229.06                        & 41.19                     & 115.08                           & 173.61                       & 484.45                          & 661.19                       &  \\
		lottery          & 83.04                         & 19.11                     & 44.60                            & 73.59                        & 123.06                          & 165.33                       &  \\
		med-stats        & 198.24                        & 26.59                     & 83.14                            & 126.40                       & 360.82                          & 491.32                       &  \\
		power-grid       & 89.11                         & 14.48                     & 40.67                            & 62.02                        & 155.76                          & 206.74                       &  \\
		receipts         & 167.25                        & 34.18                     & 104.30                           & 155.31                       & 410.89                          & 558.92                       &  \\
		reviews          & 188.67                        & 36.17                     & 112.62                           & 159.67                       & 473.02                          & 654.75                       &  \\
		sum-ring         & 140.72                        & 24.83                     & 81.75                            & 119.67                       & 326.09                          & 430.86                       &  \\
		token            & 154.25                        & 24.77                     & 70.39                            & 109.05                       & 298.09                          & 410.35                       &  \\ \midrule
		Mean             & 160.89                        & 26.82                     & 80.71                            & 121.97                       & 335.31                          & 456.63                       &  \\
		Speedup   & -                          & 6.00                      & 1.99                             & 1.32                         & 0.48                            & 0.35                         &  \\ \bottomrule
	\end{tabular}
\end{table}

\paragraph{RAM Usage}
\Cref{tab:trans-peak-mem} compares the peak memory usage during scenario
execution. Like for compilation, RSA backends are very memory-intensive.
However, the memory requirements for hybrid crypto backends are moderate.

\begin{table}
	\caption{Peak memory usage during scenario execution {[}MB{]}.}
	\label{tab:trans-peak-mem}

	\centering
	\begin{tabular}{l|rr|rrrrl@{}}
		\toprule
		& \multicolumn{2}{c|}{``dummy encryption''}& \multicolumn{5}{c}{\name}\\
		Scenario         & \multicolumn{1}{l}{\oldname} & \multicolumn{1}{l|}{\name} & \multicolumn{1}{l}{ecdh-chaskey} & \multicolumn{1}{l}{ecdh-aes} & \multicolumn{1}{l}{rsa-pkcs1.5} & \multicolumn{1}{l}{rsa-oaep} &  \\ \midrule
		exam                    & 1584.79                       & 502.45                    & 2341.76                          & 3204.56                      & 11348.01                        & 17132.19                     &  \\
		income                  & 1562.11                       & 291.39                    & 1205.80                          & 3072.60                      & 9444.64                         & 9543.21                      &  \\
		insurance               & 1548.90                       & 533.97                    & 2152.11                          & 3133.64                      & 9789.28                         & 12879.11                     &  \\
		lottery                 & 1222.43                       & 278.02                    & 759.79                           & 2608.45                      & 2342.67                         & 2703.75                      &  \\
		med-stats               & 1215.16                       & 382.98                    & 1633.88                          & 2874.27                      & 10100.89                        & 11413.27                     &  \\
		power-grid              & 1861.60                       & 293.80                    & 1203.95                          & 3105.70                      & 9278.91                         & 9502.40                      &  \\
		receipts                & 1217.36                       & 313.16                    & 1261.17                          & 2954.74                      & 9266.15                         & 9434.11                      &  \\
		reviews                 & 1218.19                       & 611.86                    & 2749.75                          & 3296.79                      & 12893.10                        & 18672.79                     &  \\
		sum-ring                & 1199.20                       & 371.73                    & 1488.59                          & 3235.51                      & 9609.82                         & 9470.62                      &  \\
		token                   & 1481.12                       & 443.35                    & 2024.75                          & 3245.45                      & 9985.55                         & 13182.52                     &  \\ \midrule
		Mean                    & 1411.09                       & 402.27                    & 1682.15                          & 3073.17                      & 9405.90                         & 11393.40                     &  \\
		Reduction Factor & -                          & 3.51                      & 0.84                             & 0.46                         & 0.15                            & 0.12                         &  \\ \bottomrule
	\end{tabular}
\end{table}

\paragraph{Summary}
In \cref{fig:scenario-scaling}, we visualize the relative differences in the
scenario runtime and memory usage for the different backends in \name. All
values are normalized with respect to \oldname (which uses ``dummy
encryption''). In general, hybrid crypto backends are more efficient than RSA
backends in terms of both scenario runtime and memory usage.

\begin{figure}
	\centering
	\begin{tikzpicture}
	\begin{axis}[
	ybar,
	width=0.9\textwidth,
	height=0.6\textwidth,
	ymax=900,
	ymin=0,
	minor tick num=3,
	ymajorgrids=true,
	yminorgrids=true,
	symbolic x coords={\oldname,dummy,ecdh-chaskey,ecdh-aes,rsa-pkcs1.5,rsa-oaep},
	ylabel={Normalized Value [\%]},
	legend style={at={(0.5,-0.15)}, anchor=north, legend columns=-1},
	]

	\addplot coordinates {(\oldname, 100.00) (dummy, 16.67) (ecdh-chaskey, 50.16) (ecdh-aes, 75.81) (rsa-pkcs1.5, 208.40) (rsa-oaep, 283.81)};
	\addplot coordinates {(\oldname, 100.00) (dummy, 28.51) (ecdh-chaskey, 119.21) (ecdh-aes, 217.79) (rsa-pkcs1.5, 666.57) (rsa-oaep, 807.42)};

	\legend{Scenario Runtime, Peak Memory Usage}
	\end{axis}
	\end{tikzpicture}
	\caption{Mean scenario runtime and peak memory usage for different backends in \name, normalized w.r.t.\ \oldname.}
	\label{fig:scenario-scaling}
\end{figure}

\subsubsection{Comparison of Proving Schemes}\label{sec:eval-proving-schemes}

So far, we have analyzed \name with the GM17~\cite{gm17} proving scheme, which
is also available in \oldname. We now demonstrate how using the
Groth16~\cite{groth16} proving scheme improves zkay's performance.

In \cref{tab:cmp-ps}, we compare the average performance, memory usage and gas
costs over all scenarios for the two proving schemes using the ECDH AES crypto
backend. Groth16 is more efficient overall: While there is no significant effect
on memory usage, Groth16 decreases average compilation time and output size by
roughly 30\%, gas costs by 25\%, and scenario runtime by 20\% compared to GM17.
In \name, Groth16 is the default proving scheme.

\begin{table}
	\caption{Comparison of \name proving schemes for the ECDH AES crypto backend.}
	\label{tab:cmp-ps}

	\begin{tabular}{lrr}
	\toprule
			Metric (mean across all scenarios)                       & GM17 &
	Groth16 \\ \midrule Compilation Time {[}s{]}         & 80.93
	& 56.59                       \\
	Compilation Peak Memory {[}MB{]} & 7898.06                  & 7025.70                      \\
	Output Size {[}MB{]}             & 1558.00                  & 1081.2                      \\
	Scenario Time {[}s{]}            & 121.97                   & 100.48                      \\
	Scenario Peak Memory {[}MB{]}    & 3073.17                  & 3021.79                     \\
	Avg. Transaction Cost {[}gas{]}   & 882366.24                & 626593.67                   \\ \bottomrule
	\end{tabular}
\end{table}

\section{Usability Improvements}

\subsection{Installation}

\Name can be packaged and installed using setuptools. Package installation
automatically compiles the required libsnark interface binary from source and
makes the zkay command gobally available.

\subsection{Error Messages}

In contrast to \oldname, \name displays human-readable and descriptive error
messages. For example, the command-line interface provides precise source code
locations for type errors.

\subsection{Contract Distribution}

\Name simplifies contract distribution by providing built-in support for
bundling and exporting/importing zkay contracts along with all relevant
information such as prover keys (see \cref{sec:contract-integrity} for details).

For example, Alice can deploy and share a contract \code{contract.zkay} with Bob
as follows.

\begin{enumerate}
	\item Alice compiles and deploys \code{contract.zkay} using her local zkay compiler.
	\item Next, she bundles the locally compiled contract to an archive \code{contract.zkp} using zkay's \code{export} command.
	\item Alice then sends the archive \code{contract.zkp} to Bob using an
	off-chain channel and informs him about the address of the deployed
	contract.
	\item Bob can now import \code{contract.zkp} on his local computer using
	zkay's \code{import} command. He can then connect to and interact with
	Alice's deployed contract.
\end{enumerate}

\subsection{Configuration Files}\label{sec:conf}

\Name allows rich customization and comes with support for configuration files.

The default settings are configured in \code{config\_user.py}, any of which can
be overridden via a hierarchy of configuration files. The user may create the
following configuration files: (i)~a system-wide configuration file in
\code{\$SITE\_CONFIG\_DIR/zkay/config.json}, (ii)~a user-wide configuration file
in \code{\$USER\_CONFIG\_DIR/zkay/config.json}, and (iii)~a local configuration
file in the working directory or in a location provided using a command line
flag.~\footnote{\Name uses the appdirs library~\cite{appdirs} to determine the
location of \code{\$SITE\_CONFIG} and \code{\$USER\_CONFIG}.}

It is further possible to override any setting via a command-line parameter of
the same name, which takes precedence over all configuration files.

\subsection{Command-line Interface}

The command-line interface of \name is based on sub-commands, which are
described below. All commands support context-aware bash-autocompletion powered
by argcomplete~\cite{argcomplete}. Programmatic access to all features is
available through the the \code{zkay\_frontend} module.

\paragraph{Development}
\begin{itemize}
	\item \texttt{zkay check}: Run the type checker on the given zkay file.
	\item \texttt{zkay solify}: Strip zkay-specific features from the given zkay
	file and output the resulting Solidity code. This makes it possible to
	analyze zkay code with tools designed for Solidity.
	\item \texttt{zkay compile}: Compile the given zkay file. This includes
	type-checking, code transformation, circuit construction and compilation,
	NIZK key generation, and contract interface generation.
	\item \texttt{zkay update-solc}: Download and install the latest compatible
	version of solc.
\end{itemize}

\paragraph{Distribution}
\begin{itemize}
	\item \texttt{zkay export}: Package the necessary data (contract, manifest
	and prover keys) of the given compilation output directory as a *.zkp
	archive.
	\item \texttt{zkay import}: Import a given *.zkp archive.
\end{itemize}

\paragraph{Deployment and Interaction}
\begin{itemize}
	\item \texttt{zkay deploy-pki}: Deploy the PKI contract.
	\item \texttt{zkay deploy-crypto-libs}: Deploy library contracts required
	for the GM17 proving scheme.
	\item \texttt{zkay run}: Open an interactive shell in the context of the
	contract interface of a given compilation output directory.
	\item \texttt{zkay deploy}: Deploy the contract of a given compilation
	output directory.
	\item \texttt{zkay connect}: Connect to the contract of a given compilation
	output directory at a given address and start an interactive shell in the
	context of the contract interface.
\end{itemize}

\bibliography{bibfile}

\end{document}